\documentclass[aps,12pt,amsmath,amssymb,showpacs,tightenlines]{revtex4}
\usepackage{bm}
\usepackage{epsfig}

\def\Eqref#1{Eq.~(\ref{#1})}
\newcommand{\be}{\begin{equation}}\newcommand{\ee}{\end{equation}}
\newcommand{\bea}{\begin{eqnarray}}\newcommand{\eea}{\end{eqnarray}}
\def\negtwomu{\mskip -2mu}

\def\R{{\bm R}} \def\Rm{\R^m}
\def\pzero{p_{0}} \def\remark{\smallskip\noindent\textsf{Remark:~} }
\def\ess{$\cal S$} \def\Xph#1{X^{(#1)}}
\def\Abf{{\bm A}} \def\A{{\cal A}} \def\Ahull{\widehat\A }
\def\zerobf{{\bm0}} \def\Ebf{{\bm E}}
\def\wbf{{\bm w}} \def\superell{^{(\ell)}}
\def\lsmatrix#1#2{\left(\begin{array}{#1}#2\end{array}\right)}
\def\wee{\hbox{$\epsilon$}} \def\weewee{\hbox{$\delta$}}
\def\notwee{\hbox{$\alpha$}} \def\Abft#1{{\bm A}^{(#1)}}

\begin{document}
\title{\hfill{\small \rm Phys.\ Rev.\ E \textbf{73}, 036124 (2006)}\\~\\%
Multiple phases in stochastic dynamics: geometry and probabilities}

\author{B. Gaveau}
\affiliation{Laboratoire analyse et physique math\'ematique, 14 avenue F\'elix Faure, 75015 Paris, France}
\email{gaveau@ccr.jussieu.fr}

\author{L. S. Schulman}
\affiliation{Physics Department, Clarkson University, Potsdam, New York 13699-5820, USA} 
\email{schulman@clarkson.edu}
\affiliation{Max-Planck-Institute for the Physics of Complex Systems, N\"othnitzer Str. 38, D-01187 Dresden, Germany}
\date{\today}
\begin{abstract}
Stochastic dynamics is generated by a matrix of transition probabilities. Certain eigenvectors of this matrix provide observables, and when these are plotted in the appropriate multi-dimensional space the phases (in the sense of phase transitions) of the underlying system become manifest as extremal points. This geometrical construction, which we call an \textit{observable-representation of state space}, can allow hierarchical structure to be observed. It also provides a method for the calculation of the probability that an initial points ends in one or another asymptotic state.
\end{abstract}
\maketitle

\section{\label{sec:intro}Introduction}

In a previous publication \cite{firstorder}, we established the relation between phases, including metastable phases, and eigenvalue degeneracy, where the eigenvalues in question are in the spectrum of a matrix of transition probabilities. The context for this use of stochastic dynamics is an approach to nonequilibrium statistical mechanics based on the master equation \cite{master}. In further work, these phases played a role in defining coarse grains in statistical mechanics \cite{grains} as well as in discerning community structure in a network \cite{dynamicalmetric}. 

In the present article we re-examine the occurrence of \textit{multiple phases}. For the case of \hbox{$m\negtwomu+\negtwomu1$} phases, there turns out to be surprising simplicity in a certain $m$-dimensional space. This is a space in which the points of the state space are given coordinate values corresponding to the first $m$ observables, where by \textit{observable} we mean a slow eigenvector (in our convention, a \textit{left} eigenvector) of the transition matrix. 
Remarkably, although one might expect no particular structure to emerge from this representation, for the case of a phase transition (i.e., eigenvalue degeneracy) the points form a simplex, which is to say, there is no more than the minimal number ($m+1$) of extremal points for the convex hull of the set of state space points in this representation. We call this geometric structure the \textit{observable representation of state space}, and it provides a practical method for the computation of the probability that an arbitrary initial state reaches one or another phase. This leads in turn to potential applications far removed from statistical mechanics. Thus one can have imperfectly defined classes of final states (i.e., they are similar but not exactly the same) for a complex random walk and be able to compute probabilities for arriving at each class. In fact one need not know the classes ahead of time. Moreover, this can be done for dynamics with relatively large state spaces. For a state space of cardinality $N$, the stochastic dynamics is generated by a matrix with $N^2$ elements---which may be daunting. Nevertheless, our method requires relatively little information about that matrix: the first few eigenvalues and eigenvectors. Hence for sparse matrices, which characterize many random walks, these quantities can be computed.

A variation of the method also provides a striking diagrammatic representation of metastable phases, particularly when there is a hierarchical tree-like structure, as occurs in spin glasses. One can see shorter and shorter lived (metastable) phases peel away (in reversed time) from those that are closer to the root of the tree. In previous work \cite{hierarchical} we explored models of this sort, but in the present article there is fuller understanding and exploitation of the observables. We have also studied other features of the transition matrix spectrum, for example situations where the eigenvalues do \textit{not} drop abruptly as the index increases. This should enhance the utility of this work in the spin glass context \cite{kurchan}.

This article has two principal sections. In Sec.\ \ref{sec:extremals} we develop the mathematical basis for the assertions just made, and in Sec.\ \ref{sec:illustrations} we provide examples in which those assertions are realized. Because of the density of mathematical estimates, we begin Sec.\ \ref{sec:extremals} with an overview, which should allow an understanding of the examples without having to go through too many details. The remainder of that section is devoted to those details. Finally in Sec.\ \ref{sec:prospects} we discuss problems that may benefit from this treatment as well as mathematical issues, such as whether certain of our hypotheses might be weakened.

\section{\label{sec:extremals}Phase transitions and extremals in the space of observables}

\subsection{Overview\label{sec:overview}}

The states of the system we study are given by $x,y\in X$, and the system moves from state to state in discrete time according to transition probabilities given by a matrix $R$. For convenience in dealing with the eigenfunctions of $R$ we define it as follows
\be
R_{xy}=``\Pr\bigl(x\leftarrow y\bigr)"=\Pr\left[\hbox{State at time $(t+1)$ is}~x\mid \hbox{State at time $t$ is}~y\right]\;.
\ee
For the processes we study there is a unique stationary distribution, $p_0(x)$, which satisfies $p_0=Rp_0$, with the $p_0$ on the right of $R$. For this eigenvalue of $R$, $\lambda_0\equiv1$, the left eigenvector is simply $A_0(x)=1$, corresponding to the conservation of probability, i.e., $\sum_{x\in X} R_{xy} = 1$, because from a state $y$ you have to go \textit{somewhere}.

The eigenvalues of $R$ fall on or inside the unit circle \cite{gantmacher1, gantmacher2} and we order them by decreasing magnitude: $\lambda_0\equiv 1 \geq |\lambda_1| \ge |\lambda_2| \ge \ldots$. The corresponding right and left eigenvectors are respectively $p_k$ and $A_k$, and satisfy
\be
R p_k = \lambda_k p_k\;, \qquad A_k R = \lambda_k A_k\;,\quad k=0,1,\ldots \;. 
\label{eq:eigvecs}
\ee

Our story begins when several of $R$'s eigenvalues are nearly degenerate with $\lambda_0=1$. As we showed in \cite{firstorder}, this heralds a phase transition in the system, enabling the realization of an old dream relating eigenvalue degeneracy and phase transitions \cite{kac}. Our method of proof involved those left eigenvectors of $R$ that correspond to the slowest eigenvalues (those nearest to 1). We now find that not only was this convenient for the proof, but it also provides a graphic illustration of phase structure along with the possibility of computing auxiliary quantities such as asymptotic probabilities and time dependence.

We suppose then that $m$ of the eigenvalues (after $\lambda_0$) are very close to 1 and that $\lambda_{m+1}$ is not. We focus on the \textit{left} eigenvectors of $R$. If $X$ has $N$ states, then we can form an $m$ by $N$ array of quantities, $A_k(x)$, with $k=1,\dots,m$, and $x\in X$. Think of the $m$-tuple $(A_1(x),\dots,A_m(x))$ as an $m$-\textit{vector}, with one such vector for each $x\in X$. All $N$ of these can be plotted in $\Rm$ and we surround this set of points with a minimal convex surface, the convex hull. In general such a surface can have many extremal points. We will show, however, that because of the eigenvalue conditions, the convex hull of this particular set of points has (essentially) just $m+1$ extrema, around each of which many, many points of $X$ may cluster. These extremal points are what correspond to the phases. To see what they look like in a typical case, with $m=2$, see Fig.\ \ref{fig:extremals1}. This is a plot of $A_1(x)$ versus $A_2(x)$. The vertices of the triangle that you see are actually composed of many points (shown in more detail in Fig.\ \ref{fig:extremals2}).

There is a straightforward intuitive way to understand this bunching. The phases are in a sense dynamically far from one another. That is, if you start in one phase you expect to stay there for a long while before going to any other phase. This means that there is a restricted dynamics within that phase that \textit{nearly} conserves probability. The bunching of points in one phase (as we define it) means that for all points in that phase $A_k(x)$ has very nearly the same value (for every $k=1,\dots,m$). Let's see why that happens. Consider the eigenvalue equation for $A_k$ applied $t$ times, where $t$ is small enough so $\lambda_k^t$ is still close to one, so close that we will now treat it as unity. Then
\be
A_k(y)\approx \sum_x A_k(x)R^t_{xy} \,.
\label{eq:Aapprox}
\ee
Now restrict $x$ and $y$ to be such that $R^t_{xy}$ is not small, so we would say $x$ and $y$ are in the same phase. Then for this restriction of $R$, \Eqref{eq:Aapprox} still holds, and we appear to have a number of eigenvectors of eigenvalue close unity. But we also have $1\approx \sum_x R^t_{xy}$, because little probability escapes the phase. This last expression says that a constant on the phase plays the role of $A_0$ for the restricted time evolution. If we now make the further assumption that relaxation \textit{within} the phase is relatively rapid, then other eigenvalues of the restricted $R$ are significantly smaller than 1, and all the apparent eigenvectors $A_k$, as well as the constant pseudo-$A_0$. must in fact be proportional to one another. In other words, the $A_k$'s ($k\leq m$) are constant on the phases. 

The actual proof proceeds a bit differently. Its heart is \Eqref{eq:9.7}, which is essentially a statement of the eigenvector property of the $A$'s. In this equation, we write $p_y^t(x)$ in place of $R^t_{xy}$, since (as just observed) the latter is the probability that starting from $y$ you reach $x$ in $t$ time steps.

\subsection{Additional properties of the stochastic matrix\label{sec:additionalproperties}}

The matrix $R$ is assumed to be irreducible. This implies that the eigenvalue 1 is unique and that its eigenvector, $\pzero$ (the stationary distribution), is strictly positive:
\be
\lambda_1\neq 1  \hbox{~~~~and~~~}
   \sum_{y\in X} R_{xy} \pzero(y) = \pzero(x)>0 \,, \quad \forall x \;. 
\ee
Since no detailed balance assumptions are made for $R$, it need not be diagonalizable nor have a spectral representation in terms of eigenvectors. Nevertheless, we will assume that for the eigenvalues that concern us (those near 1) each eigenvalue possesses one or more eigenvectors.  The orthonormality condition for the eigenvectors, $\langle A_k | p_\ell \rangle = \delta_{k \ell}$, still leaves a single multiplicative factor for each pair $(A_k,p_k)$. The stationary state $\pzero$ is naturally normalized by $\sum \pzero(x) = 1$, which fixes $A_0(x) = 1$ for all $x$. For the other $A$'s, consistent with $A_0$, we normalize by the condition
\be
\max_x |A_k(x)|=1\,,\qquad \forall k  \,.
\label{eq:normA}
\ee

Our principal assumption is that for some integer $m$, $\lambda_1, \lambda_2,\dots, \lambda_m$ are real and close to $\lambda_0\equiv1$. Specifically, this closeness is taken to mean that there exists a range of $t$ (integer times) such that for some $\epsilon\ll1$
\be
1-\lambda_k^t=\hbox{O}(\epsilon)\,,\qquad 1\leq k\leq m \,.
\label{eq:bigeigs}
\ee 
In most of our development we further assume that the $|\lambda_k|$ for $k>m$ are much smaller than $\lambda_m$, that is,
\be
|\lambda_k^t| \ll 1 \,, \qquad k\geq m+1 \,.
\label{eq:littleeigs}
\ee

\remark\ If $R$ has eigenvalues near $-1$ (or other roots of unity), our arguments go through for  $R^2$ (or other appropriate power of $R$, provided that that power is not so high that \Eqref{eq:bigeigs} is violated). See \cite{creation}, Sec.~8. 

The spectral decomposition of $R$ is written
\be
R^t=|\pzero\rangle\langle A_0|+\sum_{k=1}^{m}\lambda_k^t |p_k\rangle\langle A_k|
        +|\lambda_{m+1}^t|\, B^{(t)}
\label{eq:spectral}
\ee
with 
\be
B^{(t)}=\sum_{k\geq m}\frac{\lambda_k^t}{|\lambda_{m+1}|^t} |p_k\rangle\langle A_k|
\label{eq:dust}
\ee
\cite{note:dustjordan}.
We assume that $\lambda_{m+1}^tB$ is uniformly small in the sense that for any subset $Y\subset X$ and any $x\in X$
\be
|\lambda_{m+1}^t| \sum_{y\in Y} |B_{yx}^{(t)}| =\hbox{O}(\eta)\ll \hbox{O}(1) \,.
\label{eq:dustinequality}
\ee

\remark\ Instead of the stochastic matrix ``$R$'' we sometimes use an alternative matrix, ``$W$,'' which can be thought of as the generator of continuous time stochastic evolution. Schematically $W=(R-1)/\Delta t$. $W$ will be called a \textit{stochastic generator}. In fact it generates the usual Master Equation. The constraints on $W$ are that its off-diagonal matrix elements must be non-negative and that its column sums must vanish. Its spectrum consists of 0 and points to the left of the $y$-axis in the complex plane. Eigenfunctions are unchanged from the $R$ representation (since the matrices differ only by a multiple of the identity). The advantage of using $W$ is that in producing randomly generated matrices one need not be concerned that the column sum of off-diagonal terms be less than one.

\subsection{The phases\label{sec:phases}}

Under the foregoing hypotheses and under a separation hypothesis ``\ess'' to be introduced below, we will construct $m+1$ subsets, ultimately to be identified as the phases, $\Xph{1}, \dots, \Xph{m+1}$, of $X$ with the following properties

\begin{enumerate}
\item The sets, $\Xph{1}, \dots, \Xph{m+1}$, are disjoint.
\item On each $\Xph{\ell}$, the $A_k$ are nearly constant, for $1\leq\ell\leq m+1$ and $1\leq k \leq m$. 

\item The complementary subset of $\cup_{j=1}^{m+1} \Xph{j}$ has small $\pzero$-weight. Specifically
\be
\sum_{y\in\cup\Xph{j}}\pzero(y)\geq 1-\frac {m\epsilon}\delta-\hbox{O}(\eta) \,,
\label{eq:leftovers}
\ee
for a constant, $\delta$ such that $\epsilon/\delta$ is small (and $\epsilon\equiv1-\lambda_m^t$).

\item The $\Xph{1}, \dots, \Xph{m+1}$ are essentially unique in a sense to be described below.
\end{enumerate}

\remark\ Our phases are a bit larger than what are conventionally called phases and include states that rapidly transit to the usual phases. See Sec.~\ref{sec:basins}.

\subsection{\label{sec:existence}Proof of the existence of the phases}

For any pair $x,y\in X$, define
\be
p^t_y (x) = R^t_{xy} \,.
\label{eq:9.2}
\ee
$p^t_y(x)$ is the probability that a system in state $y$ at time 0 is in $x$ at time $t$.

Let $m\leq N-1$. Consider the following geometric construction in $\R^m$: for any $y\in X$, form the vector $\Abf(y)\equiv\left(A_1(y),\ldots , A_m(y)\right) \in \R^m$. This gives a set $\A$ of $N$ vectors in $\R^m$. Let $\Ahull$ be the convex hull of~$\A$.

\noindent
{\it The first remark is that the vector $\zerobf\equiv(0, 0,\ldots, 0)$ is in $\Ahull$.}

This follows from the orthogonality relation, $\langle  A_k | p_0 \rangle  = 0$, for $k\ge 1$. Thus
\be
\sum_{y\in X} p_0(y) \Abf(y) = \zerobf\,,
\label{eq:unlabeled1}
\ee
so that ${\bm0}$ is a convex combination of the vectors of $\A$. 

As a consequence, one can find $m+1$ points $y^*_\ell$, $1 \le \ell \le m+1$, such that the vectors
\be
\Ebf_\ell\equiv\Abf(y^*_\ell)
\label{eq:replace9.3}
\ee
are extremal points of $\Ahull$, and such that $\zerobf$ is a convex combination of them \cite{rockafellar, moulin}. There may be several ways to choose these points $y^*_\ell$, but we shall prove that, in fact, the resulting vectors $\Ebf_\ell$ ($1 \le \ell \le m+1$) are uniquely defined up to a small ambiguity to be stated later. By the selection of the $\Ebf_\ell$, we can find $\mu_\ell$, $1 \le \ell \le m+1$, with $0 \le \mu_\ell \le 1$, $\sum_{\ell =1}^{m+1}\mu_\ell = 1$ such that
\be
\Sigma \mu_\ell \Ebf_\ell = \zerobf \,.
\label{eq:9.4}
\ee

We have found that there are subsets of these points that are separated from one another in a particular way, and we add to our assumptions the following ``separation'' hypothesis (``\ess''):\\
\textsf{Hypothesis~ \ess}:~ 
For each $\ell$ let 
\be
\phi_\ell=\min_{k=1,\dots,m,\ k\neq \ell}\|\Ebf_\ell-\Ebf_k\| \,,
\label{eq:aelldef}
\ee
and define
\be
\Phi\equiv   \min_\ell     \frac{\phi_\ell}{\|\Ebf_\ell\|}  \,.
\label{eq:deltaelldef}
\ee
Then our hypothesis is that the extrema $y^*_\ell$ can be selected so that (in addition to \Eqref{eq:9.4}) they satisfy the following:
\be
 1-\lambda_m^t=\epsilon\ll \Phi \leq \hbox{O}(1)  \,.
\label{eq:hypothS}
\ee
(for an appropriate range of $t$).

We next observe that by definition
\be
\lambda^t_k A_k (y^*_\ell) = \langle  A_k | p^t_{y^*_\ell} \rangle \,,
\label{eq:9.5}
\ee
so that for all $\ell$
\be
\left( \lambda^t_1 A_1 (y^*_\ell), \lambda^t_2 A_2 (y^*_
                  \ell),\ldots, \lambda^t_m A_m (y^*_\ell)\right)
 = \sum_{y\in X} p^t_{y^*_\ell}(y) \Abf(y)
\label{eq:9.6}
\ee

Because $0 \le \lambda_k < 1$, $k\leq m$, the vector on the left side of \Eqref{eq:9.6} is in the convex set $\Ahull \,$; moreover, its distance from the extremal vector $\Ebf_\ell = \left(A_1 (y^*_\ell), \allowbreak \ldots, \allowbreak A_m (y^*_\ell)\right)$ is less than $1 - \lambda^t_m$. Since $\sum_y p^t_{y^*_\ell} (y) = 1$, we have from \Eqref{eq:9.6}
\bea
\Ebf_\ell &-& \left(\lambda^t_1 A_1 (y^*_\ell), \lambda^t_2 A_2 (y^*_\ell),
      \ldots, \lambda^t_m A_m (y^*_\ell)\right)  \nonumber\\
 && = \sum_y p^t_{y^*_\ell} (y)\bigl(\Ebf_\ell - \Abf(y)\bigr)
\label{eq:9.7}
\eea

The idea of the next few steps is as follows. For the case $m=1$ the above equation immediately yields the desired result. On the left you have something that is very small. On the right a sum of products, each factor of which is positive. Therefore one or the other of these factors must be small. This means that if $y$ can be reached from $y^*$ (with moderate probability), then $A_1(y)$ cannot be much different from $A_1(y^*)$ (in this case $\Ebf_1$ is just $A_1(y^*)$). For $m>1$ the positivity of $\bigl(\Ebf_\ell - \Abf(y)\bigr)$ is not manifest, so that it is useful to change coordinates and take as origin the vertex of the cone based on $\Ebf_\ell$. In this way all distances away from the vertex are positive in an appropriate coordinate system. \Eqref{eq:9.7} then yields the near constancy of the $A$'s on the phase.

We now pick a particular $\ell$, and consider the translated set $\Ebf_\ell - \A$ in $\R^m$, i.e., the set of vectors $\{\Ebf_\ell - \Abf(y)\mid y \in X\}$. Then, the vector $\zerobf$ is extremal for the convex set $\Ebf_\ell - \Ahull$. We prove the following lemma:

\medskip
\noindent\textsf{Lemma\/}: Consider a finite set of vectors $\wbf(y)\in\Rm$, $y\in X$, 
such that $\zerobf$ is an extremal point of the convex hull of those vectors.  Let 
$\wbf_0\in\Rm$ be a vector in the convex hull of the $\{\wbf(y)\}$. Thus
\be
\wbf_0= \sum_{y\in X} q(y) \wbf(y) \,,
\ee
with $0 \le q(y) \le 1$, $\sum q(y) = 1$. We further assume that $\|\wbf_0\|<\rho$ for some positive $\rho$, where the norm is the maximum norm, as in \Eqref{eq:normA}. Then,\\
\smallskip\noindent
(i) There exist $m$ independent linear forms on $\Rm$, $h_1,\dots,h_m$, such that $h_j (\wbf(y)) \geq 0$, $h_j(\wbf_0)<\rho$. Moreover, $h_j (\wbf(y))=0$, $\forall j$, if and only if $y$ is in the nonempty set, $\{y\}$, associated with the extremal point $\zerobf$. These forms can be scaled so as to give the same distances that we now use in $\Rm$.\\
\smallskip
\noindent
(ii) Let $a$ be a strictly positive real number. Define
\be
X(a) = \left\{y \in X  
              \,|\, 
                         h_j (\wbf(y)) < a \hbox{ for all } j = 1, \ldots, m\right\}
\,.\label{eq:9.9}
\ee
Then 
\be
\sum_{y\not\in X(a)} q(y) < \frac{m\,\rho}{a} 
\,.\label{eq:qbound}
\ee

\smallskip
\noindent
Proof: Assertion (i) comes from the fact that zero is an extremal point of the convex hull of the vectors $\wbf(y)$ for $y\in X$. The linear forms are essentially a local coordinate system with the extremal at the vertex. Positive coordinates, not necessarily orthogonal can be defined for this cone. Since the scale of this coordinate system in general includes an arbitrary multiplicative constant, it can be taken to be the same as that of the original $\Rm$. Therefore the coordinates of $\wbf_0$ in this system remain of order $\rho$.

\smallskip
\noindent
Assertion (ii): One has, by the hypotheses of the lemma, 
$$
\rho > h_j(\wbf_0) = \sum_{y\in X} q(y) h_j(\wbf(y))
   >  \sum_{\{y\in X \,|\, h_j(\wbf(y))>a\}} a q(y) \,.
$$
Thus, 
\be
\sum_{y\not\in X(a)} q(y)
 \le \sum_{j=1}^m \left(\sum_{\{y\in X \,|\, h_j(\wbf(y))>a\}}  q(y) \right)
 \le \frac{m\,\rho}a  \,.
\label{eq:proboutside}
\ee
This proves the lemma.

\remark\ In practice, the ``$m$'' appearing in \Eqref{eq:proboutside} may be a severe overestimate. This is because points in other phases will generally exceed $a$ for \textit{all} components of the linear forms. 

\medskip\noindent
\textsf{Application of the lemma}:  For every $\ell$, $1\leq \ell\leq m+1$, we apply the lemma to the set of vectors and numbers
\bea
\wbf(y)&=&\Ebf_\ell-\Abf(y)\\
\wbf_0 &=&\Ebf_\ell - \left( \lambda^t_1 A_1 (y^*_\ell), 
          \lambda^t_2 A_2 (y^*_\ell),\ldots, \lambda^t_m A_m (y^*_\ell)\right)\\
q(y) &=&   p^t_{y^*_\ell} (y)  \label{eq:qdef2} ~~~~~~~~~~~~~~~~~\\
\rho &=& (1 - \lambda^t_m)\|\Ebf_\ell\|  
  \,.
\eea
These vectors and numbers satisfy the hypotheses of the lemma, because $\zerobf$ is extremal for $\Ebf_\ell-\Ahull$ (see \Eqref{eq:9.7}), $\|w_0\|<1-\lambda_m^t$, and $\|\Ebf_\ell\|\leq1$. By the lemma, we can define for every $\ell$, $1\leq \ell\leq m+1$, $m$ independent linear forms $h_{\ell,j}$, $j=1,\dots,m$, such that if we define 
\be
X_\ell(a_\ell) = \left\{y \in X
              \,|\, 
                         h_{\ell,j} (\Ebf_\ell-\Abf(y)) < a_\ell ,\; j = 1, \ldots, m\right\} \,,
\label{eq:9.9a}
\ee
for a positive real number, $a_\ell$, then from \Eqref{eq:9.9a} and the lemma
\be
\sum_{y\in X_\ell(a_\ell)} p^t_{y^*_\ell} (y) 
     \ge 1 - \frac{m}{a_\ell} \left(1 - \lambda^t_m\right)\|\Ebf_\ell\|  
     \equiv      1 - \frac{m}{\delta_\ell} \left(1 - \lambda^t_m\right)
      \,,
\label{eq:probinside}
\ee
where $\delta_\ell=a_\ell/\|\Ebf_\ell\|$.

We now make use of hypothesis \ess. Because $\epsilon\ll\Phi$ we can find $\delta_\ell$ such that $\epsilon\ll\delta_\ell\ll\Phi$. This allows the probability of remaining \textit{within} a phase to be large (as in \Eqref{eq:probinside}), while maintaining near constancy of the $A_k$'s on that phase. Specifically, from \Eqref{eq:9.9a} we have
\be
\left|A_k(x)-A_k(x')\right|<a_\ell =\delta_\ell\|\Ebf_\ell\|
\ll \Phi \|\Ebf_\ell\|\leq \Phi \leq \hbox{O}(1) ,\quad \hbox{with~} x,x'\in \Xph{\ell} \,.
\label{eq:constA}
\ee

We next establish that these phases nearly exhaust $X$ in the sense of the probability measure $p_0$. The spectral decomposition of $R$ gives 
\be
p^t_{y_\ell^*}(y)=
R^t_{y\, y_\ell^*}=p_0(y)+
    \sum_{k=1}^{m}\lambda_k^t p_k(y) A_k(y_\ell^*)
        +|\lambda_{m+1}^t| B^{(t)}_{y\, y_\ell^*}
\label{eq:morespectral}
\ee
But
\be
\sum \mu_\ell \Ebf_\ell = \zerobf \,,
\label{eq:9.4a}
\ee
with $0 \le \mu_\ell \le 1$, $\sum_{\ell =1}^{m+1}\mu_\ell = 1$. By the definition of $\Ebf_\ell$ this means
\be
\sum_{\ell =1}^{m+1}\mu_\ell A_k(y_\ell^*) = 0 \;, \quad 1\leq k\leq m \,.
\label{eq:sumofA}
\ee
We next sum \Eqref{eq:morespectral} for each $\ell$ with coefficient $\mu_\ell$ to give
\be
\sum_{\ell =1}^{m+1}\mu_\ell\, p^t_{y_\ell^*}(y)
     =p_0(y)+ 0
        +|\lambda_{m+1}^t| \sum_{\ell =1}^{m+1}\mu_\ell  B^{(t)}_{y\, y_\ell^*}
     =p_0(y)+\hbox{O}(\eta) \,.
\label{eq:sumofdistAKA2}
\ee
Now sum over $y\in \cup_k X_k(a)$, i.e., all $y$ \textit{in} the phases, and interchange sums,
\be
\sum_{\ell =1}^{m+1}\mu_\ell \sum_{y\in \cup_k X_k(a_k)} p^t_{y_\ell^*}(y)
     =   \sum_{y\in \cup_k X_k(a_k)} p_0(y)+\hbox{O}(\eta) \,.
\label{eq:star2508}
\ee
Then we deduce
\be
\sum_{y\in \cup_k X_k(a_k)} p^t_{y_\ell^*}(y)
=
\sum_{y\in X_\ell(a_\ell)} p^t_{y_\ell^*}(y)+
\sum_{y\in \cup_{k\neq\ell} X_k(a_k)} p^t_{y_\ell^*}(y)
\geq
\sum_{y\in X_\ell(a_\ell)} p^t_{y_\ell^*}(y)
\geq
1-\frac m\delta_\ell\left(1-\lambda_m^t\right) \,,
\label{eq:starstar2508}
\ee
where the last step uses \Eqref{eq:probinside}. By \Eqref{eq:star2508} and \Eqref{eq:starstar2508} (and using $\delta=\min\delta_\ell$) ,
\bea
\sum_{y\in \cup X_k}p_0(y)&=&
\sum_\ell \mu_\ell \sum_{y\in \cup X_k} p^t_{y_\ell^*}(y) +\hbox{O}(\eta)\\
&\geq&
\sum_\ell \mu_\ell \sum_{y\in  X_\ell} p^t_{y_\ell^*}(y) +\hbox{O}(\eta)\\
&\geq&
\sum_\ell \mu_\ell\left(1-\frac m\delta\left(1-\lambda_m^t  \right)  \right)+\hbox{O}(\eta)
\eea
because $\sum\mu_\ell=1$. Therefore by \Eqref{eq:proboutside}
\be
\sum_{y\in \cup_{\ell} X_\ell (a_\ell)} p_0(y)
       \ge 1 - \frac m\delta \left(1 - \lambda^t_m\right) +\hbox{O}(\eta) \,.
\label{eq:9.14AKA3}
\ee
This proves statement (\ref{eq:leftovers}) with $X^{(\ell)}=X_\ell(a_\ell)$. 

On each phase $X^{(\ell)}$, one has by definition, 
\be
0\leq h_{\ell,j}\left(\Abf(y_\ell^*)-\Abf(y)\right)<a_\ell \, \qquad j=1,\dots,m  \,.
\label{eq:lessthana}
\ee
This implies that the coordinates of the vector $\Abf(y_\ell^*)-\Abf(y)$, for $y\in X^{(\ell)}$, are O($a_\ell$) for $k=1,\dots,m$, because the $h_{\ell,j}$ are linearly independent forms. This establishes \Eqref{eq:dustinequality}.

\remark\ If ``$m$'' is large (for example where many metastable phases are present) the factor $m\epsilon$ may not be smaller than one. In that case there may not be the clean separation of phases that we discuss here.

\subsection{\label{sec:uniquenessproof}Proof of the uniqueness of the phases}
We start from \Eqref{eq:eigvecs} for $A_k$ and $R^t$,
\be
 \lambda_k^t A_k=A_k R^t\,,\quad 1\leq k\leq m \;. 
\label{eq:eigvecsrepeat}
\ee
The right hand side of this equation can be split by summing over each phase $\Xph{\ell}$ and over the set of points of $X$ outside of any phase:
\be
\lambda_k^t A_k(y)=\sum_{\ell=1}^{m+1}\left(
     \sum_{x\in\Xph{\ell}} A_k(x)R^t_{xy} \right)
+    \sum_{x\notin\cup_{\ell}\Xph{\ell}} A_k(x)R^t_{xy}
\label{eq:Abreakup}
\ee
We next estimate each sum in \Eqref{eq:Abreakup}. One has $|A_k(x)|\leq1$, by normalization. Moreover,
\be
\sum_{x\notin\cup_{\ell}\Xph{\ell}} R^t_{xy}
\le
K \frac{1-\lambda_m^t}{\delta_\ell} +\hbox{O}(\eta)  \,,
\label{eq:dustoutside}
\ee
where $K$ is O(1), as will be shown below in Sec.\ \ref{sec:weights}, and in particular \Eqref{eq:boundall}. Choose a point $x\superell\in\Xph{\ell}$ for $1\leq\ell\leq m+1$. One has 
\be
\sum_{x\in \Xph{\ell}} A_k(x)R^t_{xy} =
A_k(x^{(\ell)})\sum_{x\in \Xph{\ell}} R^t_{xy}+
\sum_{x\in \Xph{\ell}}\left( A_k(x)-A_k(x^{(\ell)})\right)R^t_{xy}
\label{eq:Abreakup2}
\ee
The last sum in \Eqref{eq:Abreakup2} is O$(a_\ell)\sum_{x\in \Xph{\ell}}R^t_{xy}$, using \Eqref{eq:constA}, which is in turn O($a_\ell$), since $\sum_{x\in \Xph{\ell}}R^t_{xy}<1$. Therefore \Eqref{eq:Abreakup} becomes
\be
A_k(y)=\sum_{\ell=1}^{m+1}\left(\frac1{\lambda_k^t}
              \sum_{x\in\Xph{\ell}}R^t_{xy}\right)A_k(x^{(\ell)}) +\hbox{O}(a_\ell)+\hbox{O}(\eta) \,,
\label{eq:Aestimate}
\ee
with $a\equiv \max_{1\leq\ell\leq m+1} a_\ell\ll\hbox{O}(1)$. Moreover, $\lambda_k^t= 1+\hbox{O}(\epsilon)$, so that finally
\be
A_k(y)=\sum_{\ell=1}^{m+1} q_\ell(y)A_k(x^{(\ell)}) +\hbox{O}(\eta)+\hbox{O}(a)
\label{eq:Aestimate2}
\ee
with
\be
q_\ell(y)=\sum_{x\in\Xph{\ell}}R^t_{xy} \,.
\label{eq:qdef}
\ee
$q_\ell(y)$ is the probability that, starting from $y\in X$, the system is in phase $\Xph\ell$ at time $t\gg1$ \cite{note:qexplanation}. The system of \Eqref{eq:Aestimate2} is a vector equation
\be
\Abf(y)
      =\sum_{\ell=1}^{m+1} q_\ell(y)\Abf\left(x^{(\ell)}\right)
           +\hbox{O}(a)
\label{eq:famousbound}
\ee
Moreover, one has
\be
\sum_{\ell=1}^{m+1} q_\ell(y)=\sum_{\ell=1}^{m+1}\sum_{x\in\Xph\ell} R^t_{xy}
    =1-\sum_{x\notin\cup\Xph\ell} R^t_{xy}=1+\hbox{O}(\epsilon)
\label{eq:qsum}
\ee
because of \Eqref{eq:dustoutside}.

Thus \Eqref{eq:famousbound} says that the vector $\Abf(y)$ for any $y$ is in the convex hull generated by the $m+1$ vectors $\Abf\left(x^{(\ell)}\right)$, up to O($a$) corrections. Therefore, up to O($a$), there can be no extremal points except those already among the $\{x^{(\ell)}\}$, since otherwise \Eqref{eq:famousbound} would be an expression for one extremal point in terms of the others. This implies that the phases are unique (up to O$(a)$).

\subsection{\label{sec:barycentric}Barycentric coordinates}

\Eqref{eq:famousbound} says that the vector $\Abf(y)$ has barycentric coordinates $q_\ell(y)$
defined by \Eqref{eq:qdef}, with respect to the vectors $\Abf\left(x^{(\ell)}\right)$ in each phase $X^{(\ell)}$, where it does not matter which $x^{(\ell)}\in X^{(\ell)}$ is chosen, up to errors of order $a$. Moreover, by \Eqref{eq:qdef}, $q_\ell(y)$ is the probability that, starting from $y$, the state of the system is in phase $X^{(\ell)}$ at time $t$ (where $t$ is the time scale used to define the phases).  

But this means that one can calculate these probabilities $q_\ell(y)$ in a geometric manner. We need to calculate the first $m$ left eigenvectors $A_1(y),\dots,A_m(y)$ (after the trivial $A_0$). 
This is sufficient to define the phases by our construction. Using these phases the $q_\ell(y)$ are the barycentric coordinates of $\Abf$ with respect to the phases. Thus, the spectral properties and convex hull construction provide a way to calculate the probability of reaching classes of intermediate asymptotic (time-scale $t$) states, that is to say, the phases.

\subsection{\label{sec:weights}Weight estimates inside and outside the phases}

We wish to establish \Eqref{eq:dustoutside}.

\paragraph*{1.}
For each $k$, $1\leq k\leq m+1$, the following relation is satisfied by the points inside the phase
\be
\sum_{x\in\Xph{k}}p^t_{y_k^*}(x)\geq 1-\frac{1-\lambda^t_m}\delta
\label{eq:weights2}
\ee
[cf.\ \Eqref{eq:proboutside}]. \newline

\paragraph*{2.}
We now consider arbitrary $x$ (not necessarily in one of the phases). By the spectral decomposition
\be
p^t_{y_k^*}(x)=
  \pzero(x)+\sum_{\ell=1}^{m} 
            \lambda_\ell^t\, p_\ell(x)A_\ell(y_k^*)+ \lambda_{m+1}^t B^{(t)}_{xy_k^*}
\,,\qquad 1\leq k\leq m+1 \,.
\label{eq:spectral2}
\ee
With the notation
\be
\alpha_{n\ell}\equiv A_{n-1}(y_\ell^*) \,,\quad p'_k\equiv p_{k-1}\,,
\quad P_j(x)\equiv p^t_{y_j^*}(x) \,,
\label{eq:alphadef}
\ee
\Eqref{eq:spectral2} becomes
\be
P_k(x)= \sum_{\ell=1}^{m+1} p'_\ell(x)\alpha_{\ell k}   +\hbox{O}(\epsilon)+\hbox{O}(\eta)
\,,\qquad 1\leq k\leq m+1 \,.
\label{eq:spectral4}
\ee
This is a linear system of $m+1$ equations for the $p_k(x)$, and as a consequence, one can solve for those quantities:
\be
p'_k(x)=
 \sum_{\ell=1}^{m+1}P_\ell(x)\, c_{\ell k}
      +\hbox{O}(\epsilon)+\hbox{O}(\eta)
\label{eq:solveforp}
\ee
with $c=\alpha^{-1}$. That is, the first $m+1$ right eigenvectors are expressible in terms of the distributions within each phase. This is the generalization of Eq.\ (3.7) of \cite{firstorder}. \\

\remark\ From this relation we can also see that each $p_k$ (of the first $m+1$) right eigenvectors is proportional to $p_0$ on each phase; that is, $p_k(x)\approx\,\hbox{const}\cdot p_0(x)$ for $\{x\}$ that constitute a single phase. First recall that each $p^t_{y_\ell^*}(x)$ looks locally like $p_0$ on its phase, because $t$ is assumed large enough that local equilibration is complete (that's the smallness of $\lambda_{m+1}^tB$). On the other hand, outside phase-$\ell$ each $p^t_{y_\ell^*}(x)$ is zero, since starting from the extremal on a time scale such that $\lambda_m^t$ is still close to unity there is little escape. Therefore, by \Eqref{eq:solveforp}, each $p_k$ (or $p'_k$) is simply given as a sum of such $p^t_{y_\ell^*}(x)$ (or $P_\ell(x)$).

\medskip

\paragraph*{3.}
Next take the points $x$ \textit{outside} all phases and write again, for any $y^\dagger$ (inside or outside of any phase)
\be
p^t_{y^\dagger}(x)=
  \pzero(x)+\sum_{\ell=1}^{m} 
            \lambda_\ell^t p_\ell(x)A_\ell(y^\dagger)+ \lambda_{m+1}^t B^{(t)}_{xy^\dagger} \,.
\label{eq:spectral3}
\ee
In \Eqref{eq:spectral3} replace the $p_k(x)$'s by their values given in \Eqref{eq:solveforp}
\be
p^t_{y^\dagger}(x)=
   \sum_{k=0}^{m}
      \sum_{\ell=1}^{m+1}
        \lambda_k^t A_k(y^\dagger)c_{k\ell}p^t_{y_\ell^\dagger}(x) + \lambda_{m+1}^t C^{(t)} \,,
\label{eq:replace}
\ee
where $C^{(t)}$ is a combination of the various $B^{(t)}$ and is assumed to be bounded. But from \Eqref{eq:probinside}
\be
\sum_{x\notin \Xph k}p^t_{y_k^*}(x) \leq \frac m\delta \left(1-\lambda^t_m\right) \,,
\label{eq:oldbound}
\ee
so from \Eqref{eq:replace}
\be
\sum_{x\notin \cup_k\Xph k}
         p^t_{y^\dagger}(x) 
           \leq K \frac{1-\lambda^t_m}\delta
      +\hbox{O}(\eta)
\,,
\label{eq:boundall}
\ee
where $K$ depends on the $c_{k\ell}$ and thus only on the geometry of the $\Abf\left(y_k^*\right)$ and is therefore O(1). Therefore at time $t$, the probability of starting from $y^\dagger$ (for any $y^\dagger$) to be outside of all phases is small, when $t$ is such that 
\be
\frac{1-\lambda^t_m}\delta \ll\hbox{O}(1) \,, 
\qquad\hbox{and}\quad 
|\lambda^t_{m+1}||B^{(t)}|\ll\hbox{O}(1) \,.
\label{eq:oneminuslambdabounds}
\ee

\paragraph*{4.}
We know that
\be
\sum_{x\notin \cup_k\Xph k} \pzero(x) \leq  \frac m\delta \left(1-\lambda^t_m\right)
      +\hbox{O}(\eta)  \,.
\label{eq:pzerobound}
\ee
For this last estimate, the left hand side does not depend on $t$, so that the right hand side can be taken a value of $t$ such that \Eqref{eq:oneminuslambdabounds} is valid.

\subsection{Basins of attraction\label{sec:basins}}

As noted above, our phases are not only the usual states in a phase, but also include the short-time-scale basins of attraction for those phases. In this context ``short-time-scale'' means O($\eta$), that is, at worst, the next slower mode after $\lambda_m$. A particular example of this can occur when the extremum point itself is not in what one usually calls the phase but only gets there in O($\eta$). In this behavior, the extremal is like the points that are not uniquely identified with any phase and have non-zero probabilities for going to several. (See below, Fig.\ \ref{fig:extremals1}, for an example of this.) The only difference between these intermediate points and a not-in-the-phase extremal is that the barycentric coordinates for such an extremal have a single 1, with the other entries zeros. The distinction between basin-of-attraction states and those conventionally assigned to the phase will lie in the size dependence of $p_0(x)$---which assumes that one is in a conventional context, where a thermodynamic limit is contemplated.

\begin{figure}
\centerline{
\includegraphics[height=.28\textheight,width=.45\textwidth]{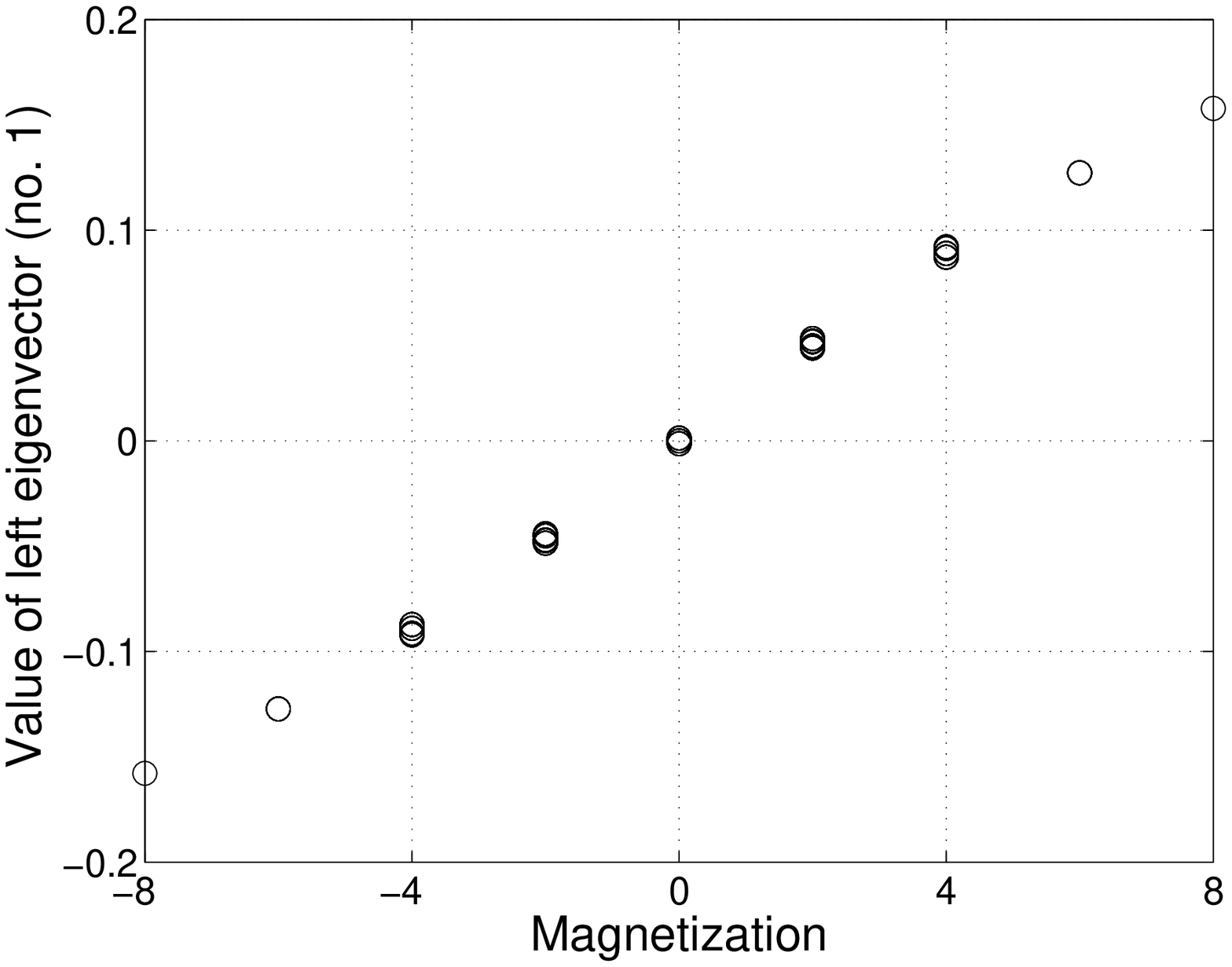}~
\includegraphics[height=.28\textheight,width=.45\textwidth]{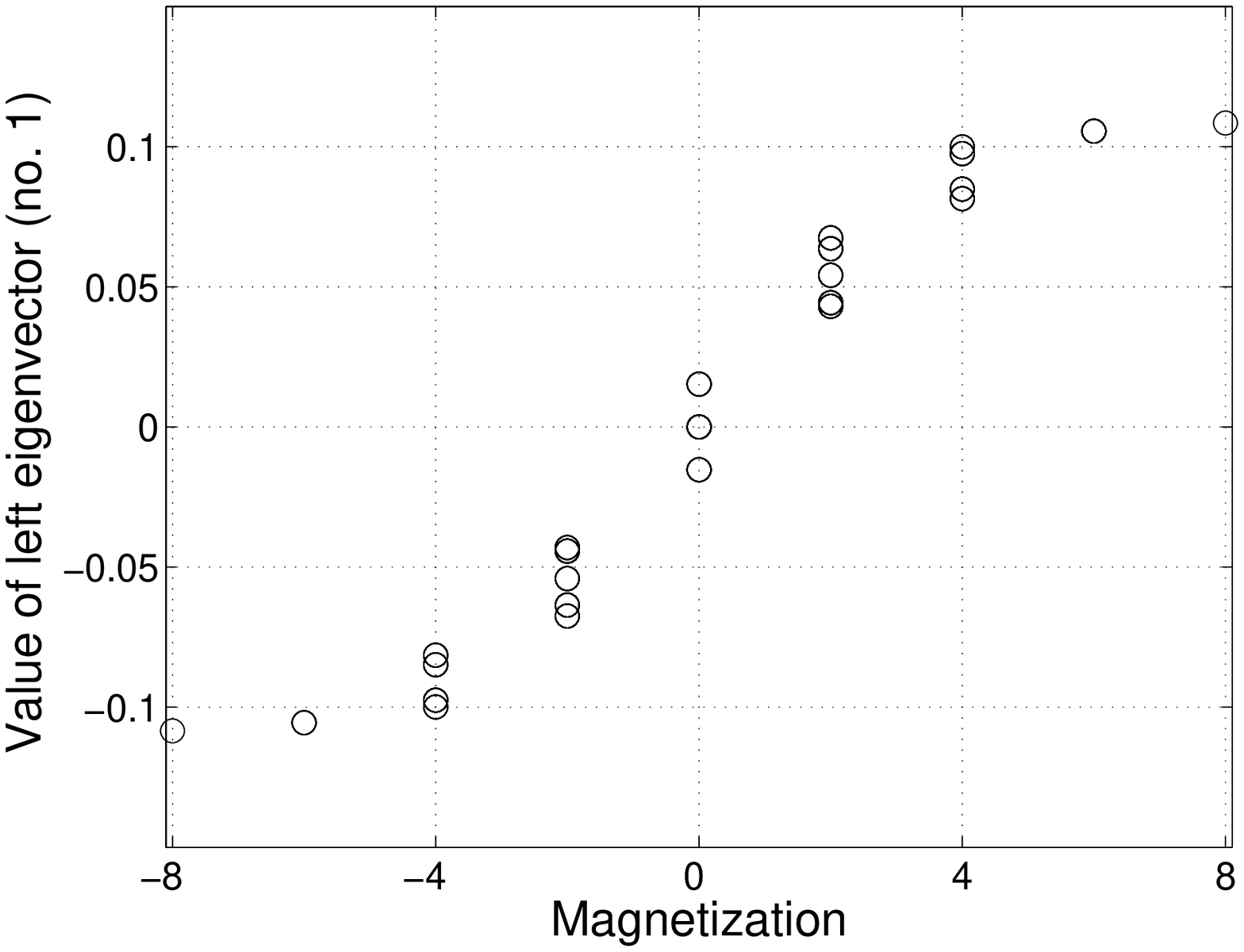}}%
\caption{Ising model stochastic dynamics. $A_1$ versus magnetization. On the left is shown $T>1$, on the right $T<1$. In this figure an L$^2$ normalization is used for $A_k$ (see \cite{hierarchical}). \label{fig:ising}}
\end{figure}

An illustration of an extremal that does not lie in the usual phase can be found in our article on the definition of coarse grains, \cite{grains}. There we analyzed a 1-dimensional Ising model. Although there is no conventional phase transition in this system, there is a marked difference in system behavior for temperature above and below the value $T=1$ (in the units we use there) even for moderate values of the number of spins on the ring. In that article we plotted the left eigenvector, $A_1(x)$, as a function of magnetization. That is, $x=(\sigma_1,\dots,\sigma_N)$ is a spin configuration, with each $\sigma_k=\pm1$, and the magnetization is $\sum_k\sigma_k$. Two plots, one below $T=1$ and one above are shown in Fig.\ \ref{fig:ising}. In \cite{grains} we made the point that the magnetization emerges from the slow eigenvalues ($\lambda$ near 1) in a natural way. Surprisingly (to us) this turned out to be more marked \textit{above} $T=1$ than below. The close relation of $A_1$ to magnetization is evident in the $T>1$ figure. Below $T=1$ there was a bunching of values near the maximum (and symmetrically about the minimum) \cite{note:proofthatAismagnetization}. The maximum of $A_1$, which we use to label the phase, occurs for the state with maximum magnetization, which, using analyticity-related definitions, is not part of the phase in the thermodynamic limit \cite{note:heuristic}. Nevertheless, this maximum of $A_1$ (for $T<1$) differs little from $A_1(x)$ for other points, $x$.

\def\proofthatAismagnetization{%
For the one-dimensional Ising model (used to generate our figure), with single-spin-flip dynamics, one can show that in the high temperature limit $MW=-2\exp(-4\beta J)M+$O$(1-\exp(-4\beta J))$, with $M$ the magnetization vector, $J$ the coupling constant and $\beta$ inverse temperature. For lower temperatures additional terms become important.}

\subsection{Detailed balance for the principal portions of $R^t$ \label{sec:detbal}}

As above, assume that the first $m+1$ eigenvectors of $R$ are real and that in the spectral expansion the rest of $R$ is small. Then $R$ itself nearly satisfies detailed balance. That is, for the truncated $R$, $J_{xy}=R^t_{xy}p_0(y)-R^t_{yx}p_0(x)$ is small. 

The truncated spectral expansion for $R$ is (cf.\ \Eqref{eq:morespectral}) $\bar R^t_{x\,y}= p_0(x)+\sum_{k=1}^{m}\lambda_k^t p_k(x) A_k(y)$, from which we have
\be
\bar J_{xy}
=\bar R^t_{xy}p_0(y)-\bar R^t_{yx}p_0(x)
=   \sum_{k=1}^{m}\lambda_k^t \left[p_k(x) A_k(y)p_0(y)-p_k(y) A_k(x)p_0(x)
\right] \,.
\label{eq:Jtruncated}
\ee

\noindent
\textsf{Case 1}: $x$ and $y$ are in the same phase. Up to O($\eta$), $A_k(y)=A_k(x)$, so that every term in the sum contains differences, $p_k(x)p_0(y)-p_k(y)p_0(x)$. However, by the Remark following \Eqref{eq:solveforp}, on any particular phase  $p_k(x)$ is proportional to $p_0(x)$, so these differences are zero.

\noindent
\textsf{Case 2}: they are not in the same phase: then $R^t_{xy}$, which is $p^t_y(x)$, is close to zero (on the order of $\epsilon$). Therefore $\bar J_{xy}$ is also zero.

The foregoing observation must be used with caution. It only asserts that $\bar J_{xy}$ is small on scale of $\eta$ or of $\epsilon$. However, for times, $\tau$, such that $1-\lambda_m^\tau$ is \textit{not} small, this conclusion does not hold.

\remark\ Although detailed balance implies that all eigenvalues of $R$ are real, the converse is not true. (Thus the mere fact that all eigenvalues in the truncated $R$ are real does not already imply detailed balance.) From numerical exploration we have indeed found that there is a correlation between $\sum_{xy} |J_{xy}|$ and $\sum |\Im{\lambda_k}|$ (for matrices generated in a certain random way); nevertheless it does happen that $\sum |\Im{\lambda_k}|=0$, while $\sum_{xy} |J_{xy}|$ is not \cite{note:counterexample}.

\def\counterexample{Here is a 3-by-3 example of a stochastic generating matrix (``$W$'' not ``$R$'') that provides a counterexample: $W_{12}=0.486$, $W_{13}=0.457$, $W_{21}=0.231$, $W_{23}=0.019$, $W_{31}=0.607$, $W_{32}=0.762 $ (with appropriate diagonal to satisfy $\sum_x W_{xy}=0$). This has eigenvalues $0$, $-1.20$ and $-1.36$, all real. The calculated positive current is 0.458 times the permutation matrix with 1 in the (1,3) position}

\subsection{A reduced stochastic process\label{sec:reduced}}

Until now we have focused on a time scale ``$t$'' such that $1-\lambda_m^t\ll1$. Now take a time, $T$, such that this is not the case. For such $T$, we can drop O($\eta$) and O($\epsilon$) terms in our representation of $R$, but nevertheless need to retain $1-\lambda_m^T$. This also implies that for $k=1,\dots,m$, $A_k(y)$ can be replaced by $A_k(y_j^*)$ for $y\in \Xph j$. Furthermore, for these longer times, points not in any phase can be dropped, since they have long before made their way to one or another  phase.

Start with the standard spectral representation (dropping O($\eta$) and O($\epsilon$))
\be
R^T_{xy}=\sum_{k=0}^m \lambda_k^T p_k(x)A_k(y_\ell^*)  \,,\quad\hbox{for~} y\in \Xph\ell          \,.
\label{eq:Rspectral}
\ee
Take $x\in\Xph k$ and do a standard coarse graining \cite{grains}:
\be
\widetilde R(k,j)=\sum_{x\in \Xph{k},\, y\in\Xph{j}}R^T_{xy}\frac{p_0(y)}{\mu(j)}
  =\sum_{\ell=1}^{m+1} \lambda_{\ell-1}^T 
         \left(\sum_{x\in \Xph{k}}p'_{\ell}(x)\right)
              \left(\sum_{y\in\Xph{j}}\alpha_{\ell j}\frac{p_0(y)}{\mu(j)}\right) \,,
\label{eq:coarse1}
\ee
where $\mu(j)\equiv \sum_{x\in\Xph{j}}p_0(x)$. (N.B. $\mu(k)$, the measure, and $\mu_k$, the barycentric coordinate, are not the same.) Recall from the observation following \Eqref{eq:solveforp} that $p'_n(x)$ is proportional to $p_0(x)$ in each phase. Define the proportionality constant by $p_\ell(x)=p_{\ell k} p_0(x)$ (or $p'_{\ell}(x)=p'_{\ell k} p_0(x)$). Doing the sums, \Eqref{eq:coarse1} becomes
\be
\widetilde R(k,j)=\sum_{\ell=1}^{m+1} \lambda_{\ell-1}^T 
         p'_{\ell k}\mu(k) 
              \left(\alpha_{\ell j}\frac{\mu(j)}{\mu(j)}\right) 
              =\sum_{\ell=1}^{m+1} \lambda_{\ell-1}^T 
                 p'_{\ell k}\mu(k)
                   \alpha_{\ell j} \,,
\label{eq:coarse2}
\ee
The fact that $\sum_j\widetilde R(k,j)=1$ follows immediately by summing over $x$ in \Eqref{eq:Rspectral}.

The stochastic matrix $\widetilde R$ therefore describes the transitions between phases on a much-elongated time scale.

\section{Illustrations\label{sec:illustrations}}

In this section we illustrate the general principles with specific numerical examples.
\subsection{Multiple phases with relatively rapid internal relaxation\label{sec:multiple}}

\begin{figure}
\includegraphics[height=.4\textheight,width=.7\textwidth]{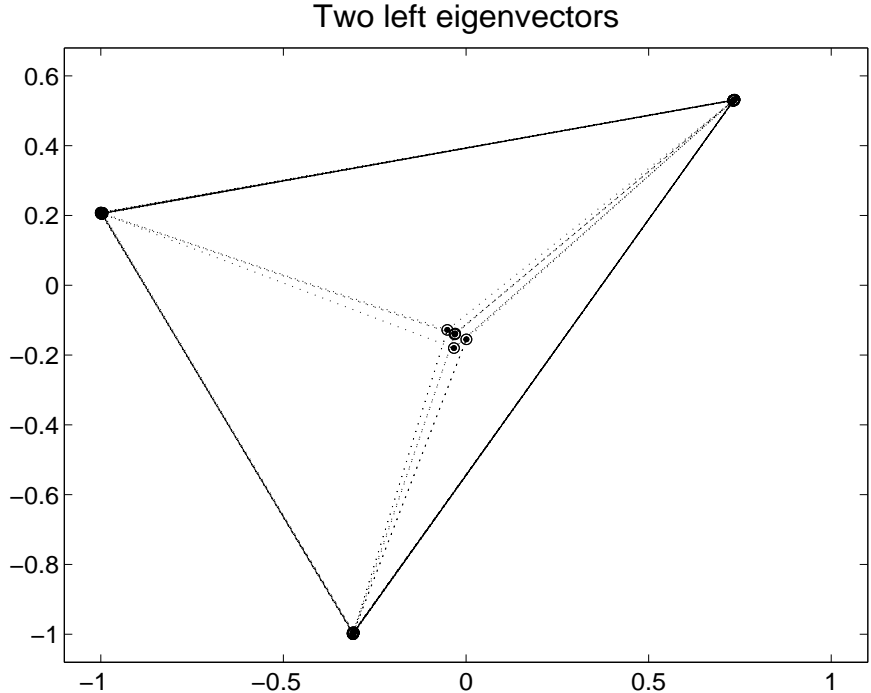}%
\caption{Plot using the first two left eigenvectors ($A_1$ and $A_2$) of the transition matrix, $R$, for a three-phase system. A circle is placed at each point $(A_1(x),A_2(x))$ for each of the $N$ states, $x$, in $X$. The lines connecting the circles are for visualization. The matrix $R$ is generated by combining 4 blocks, 3 of which are random matrices, the fourth essentially zero. Then a bit of noise is added throughout, with bigger terms for migration out of the fourth block. Finally the diagonal is adjusted to make the matrix stochastic. This leads to a pair of eigenvalues near one. This plot using the first two eigenvectors shows the extremal points to be clustering in three regions, corresponding to the phases. The points not at the extremals represent the fourth block, all of which head toward one or another phase under the dynamics. For the particular matrix chosen, they are about as likely to end in one phase as another. For all eigenvector plots, the quantities plotted are pure numbers whose scale is set by our normalization convention, discussed in Sec.\ \ref{sec:additionalproperties}.}
\label{fig:extremals1}
\end{figure}

In Fig.\ \ref{fig:extremals1} we show a three phase situation. 

For simplicity we work with the matrix ``$W$'', discussed in the Remark just before Sec.~\ref{sec:phases}. The form of the matrix corresponding to this figure is schematically
\be
\widetilde W=\lsmatrix{cccc}{
W_1&\wee&\wee&\notwee\\
\wee&W_2&\wee&\notwee\\
\wee&\wee&W_3&\notwee\\
\wee&\wee&\wee& 0
}
+\wee\cdot\hbox{random} \,,\qquad W=\widetilde W - \hbox{diag}\left(\sum \widetilde W\right) \,.
\label{eq:3phaseW}
\ee
Thus 3 random matrices are produced and weakly coupled to one-another: ``$\epsilon$'' is a generic small matrix and need not be the same matrix for each appearance in $\widetilde W$. Then additional states are added to the state space (those appearing after $W_3$). These have large one-way couplings to the other states (``$\alpha$'', a generic not-small matrix which again need not be the same in each of its appearances in $\widetilde W$) plus small probabilities of return. Next small transition probabilities are added to be sure the matrix is ergodic. Finally the actual stochastic generating matrix, $W$, is computed from $\widetilde W$ by forming column sums and subtracting those sums on the diagonal (where for $A$ an $n\times n$ matrix, $\sum A$ is the $n$-component object $\sum_x A_{xy}$ and ``diag'' is a diagonal $n\times n$ matrix with its argument on the diagonal and zeros elsewhere).

\begin{figure}
\includegraphics[height=.35\textheight,width=.6\textwidth]{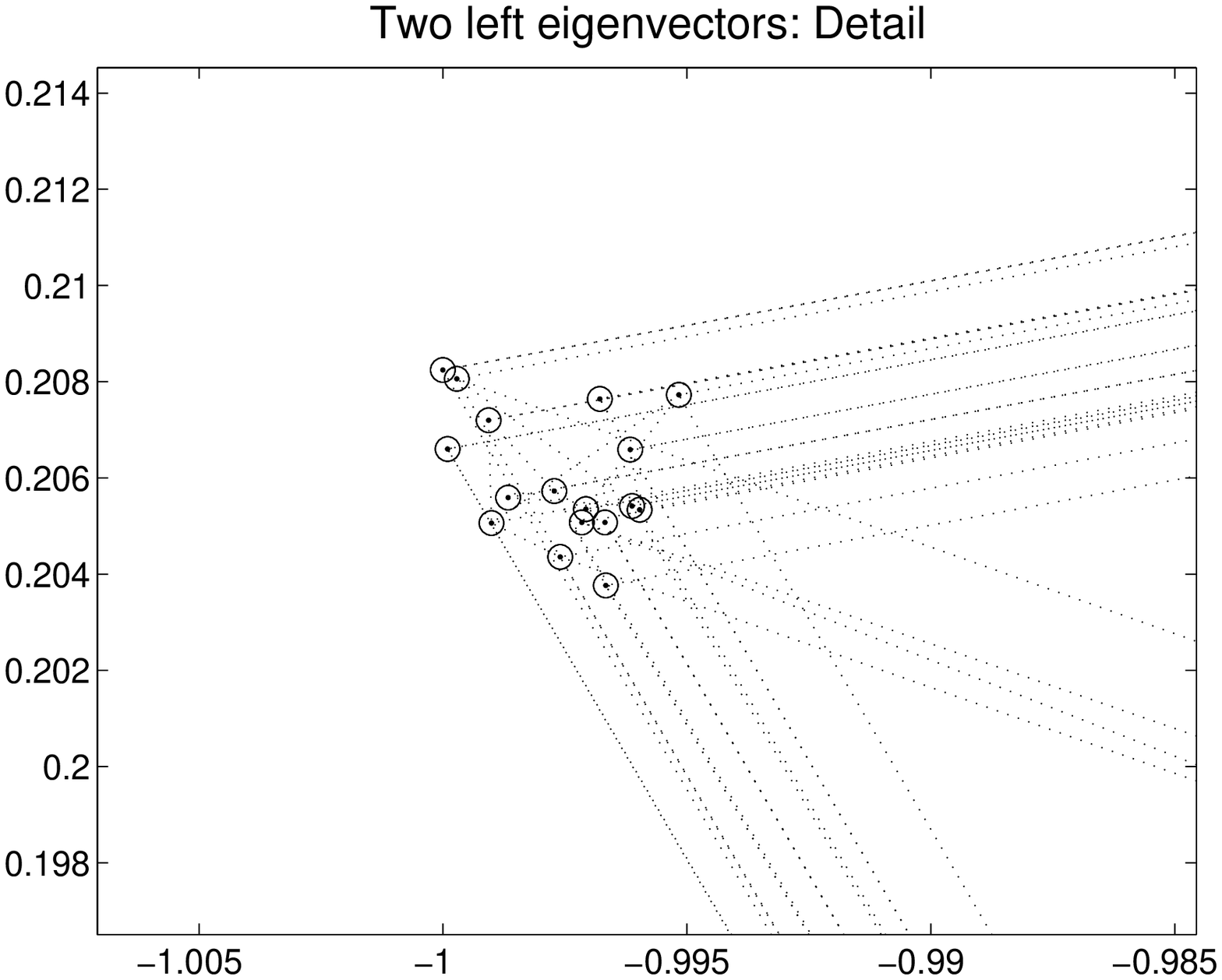}%
\caption{Detail of the upper left vertex in Fig.\ \ref{fig:extremals1}. In actuality the points in each phase cluster closely together and more than one extremal might, in principle, occur. The precision is limited by the non-negligible magnitudes of the quantities $1-\lambda_2$ and $\lambda_3$.}
\label{fig:extremals2}
\end{figure}

Because this is a three-phase system (by construction) it is sufficient to take $\Abf(x)$ to be 2-dimensional, i.e., we plot only $A_1(x)$ versus $A_2(x)$ for all $x\in X$. This is Fig.\ \ref{fig:extremals1}. The vertices of the triangle consist of the large number of points (dimensions of the spaces associated with $W_k$, $k=1,2,3$), while near the middle of the triangle are the points associated with the additional dimensions in the lower right hand corner of $W$. Because the form of $\alpha$ was approximately the same for all of them, they are near one another. If any of these points is expressed in barycentric coordinates with respect to the vertices, the coefficients give the probability that starting from this point one reaches the respective phase (vertex). To see that the vertices are actually blurred a bit, we plot in Fig.\ \ref{fig:extremals2} a close-up of one of the vertices. This is the \textit{same} matrix as in Fig.~\ref{fig:extremals1}.

\begin{figure}
\includegraphics[height=.3\textheight,width=.6\textwidth]{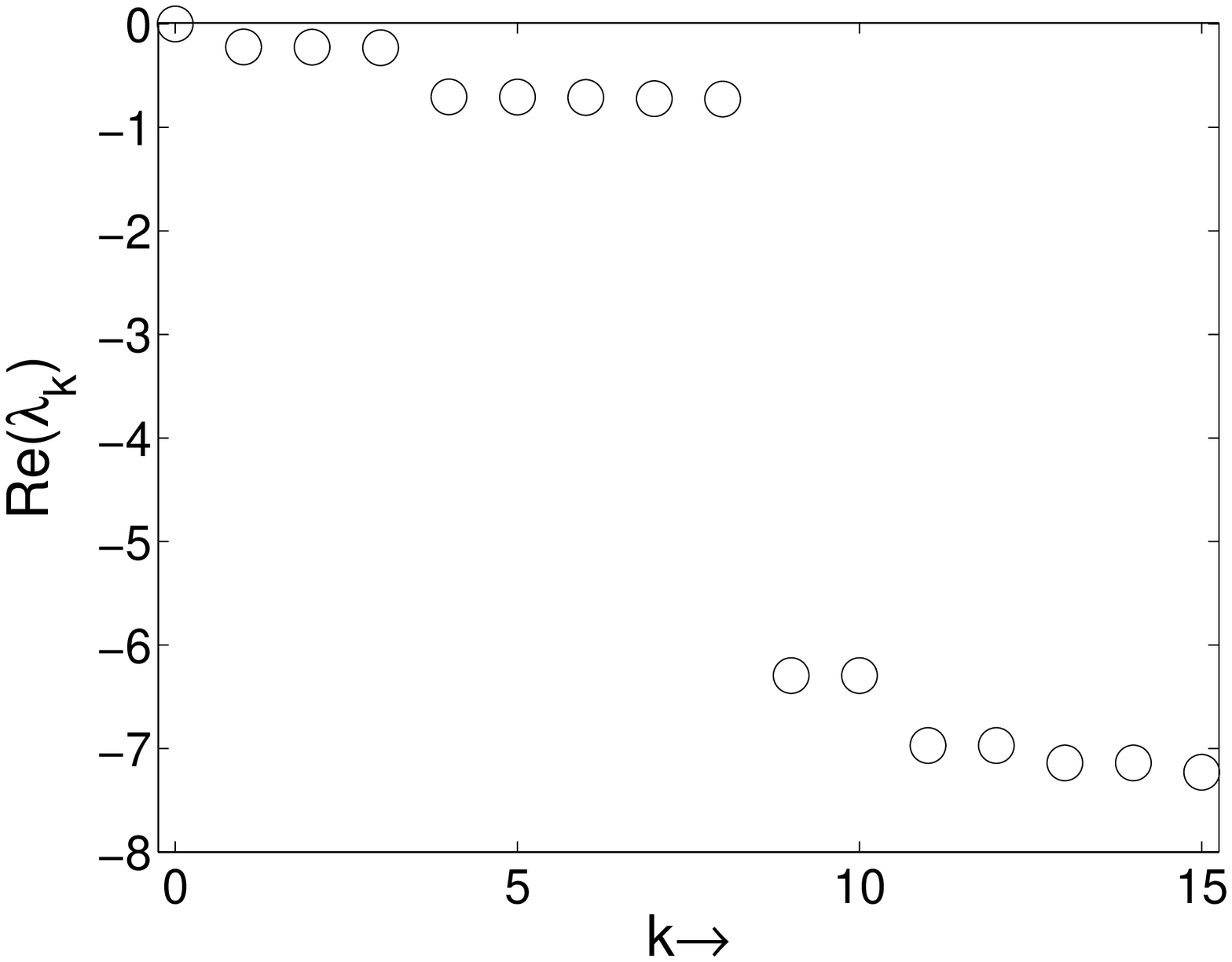}
\caption{The first few eigenvalues of $W$ for the four phase system. For those eigenvalues having an imaginary part (which is not the case for the first four), only the real part is shown. }
\label{fig:extremals3}
\end{figure}

\begin{figure}
\includegraphics[height=.4\textheight,width=.7\textwidth]{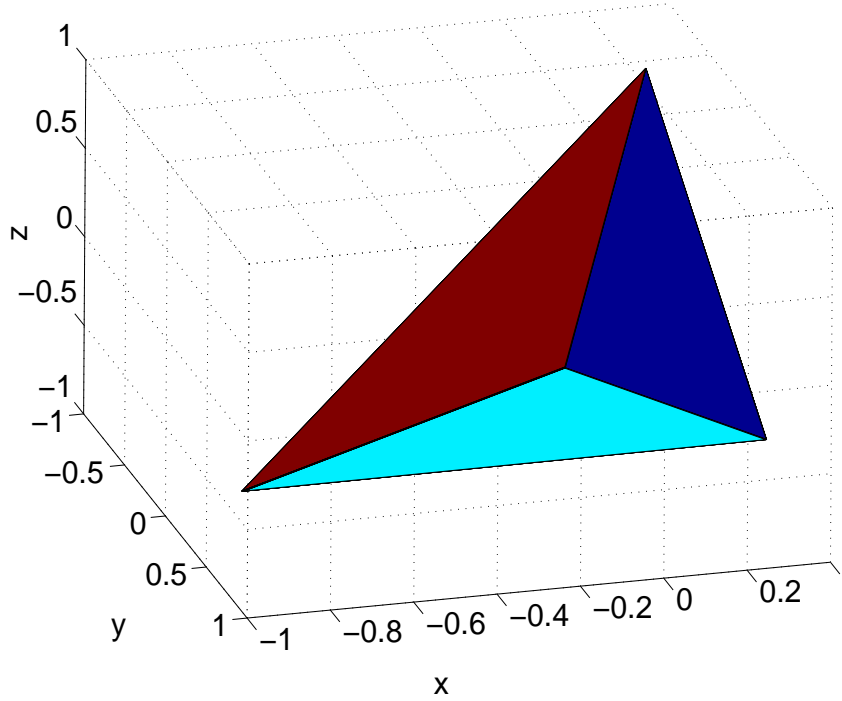}
\caption{(Color online) Convex hull of the set of points $\Abf(y)$ for $y\in X$. This is for a case of 4 phases and the figure formed in $\R^3$ is a tetrahedron.}
\label{fig:extremals4}
\end{figure}

\begin{figure}
\includegraphics[height=.4\textheight,width=.7\textwidth]{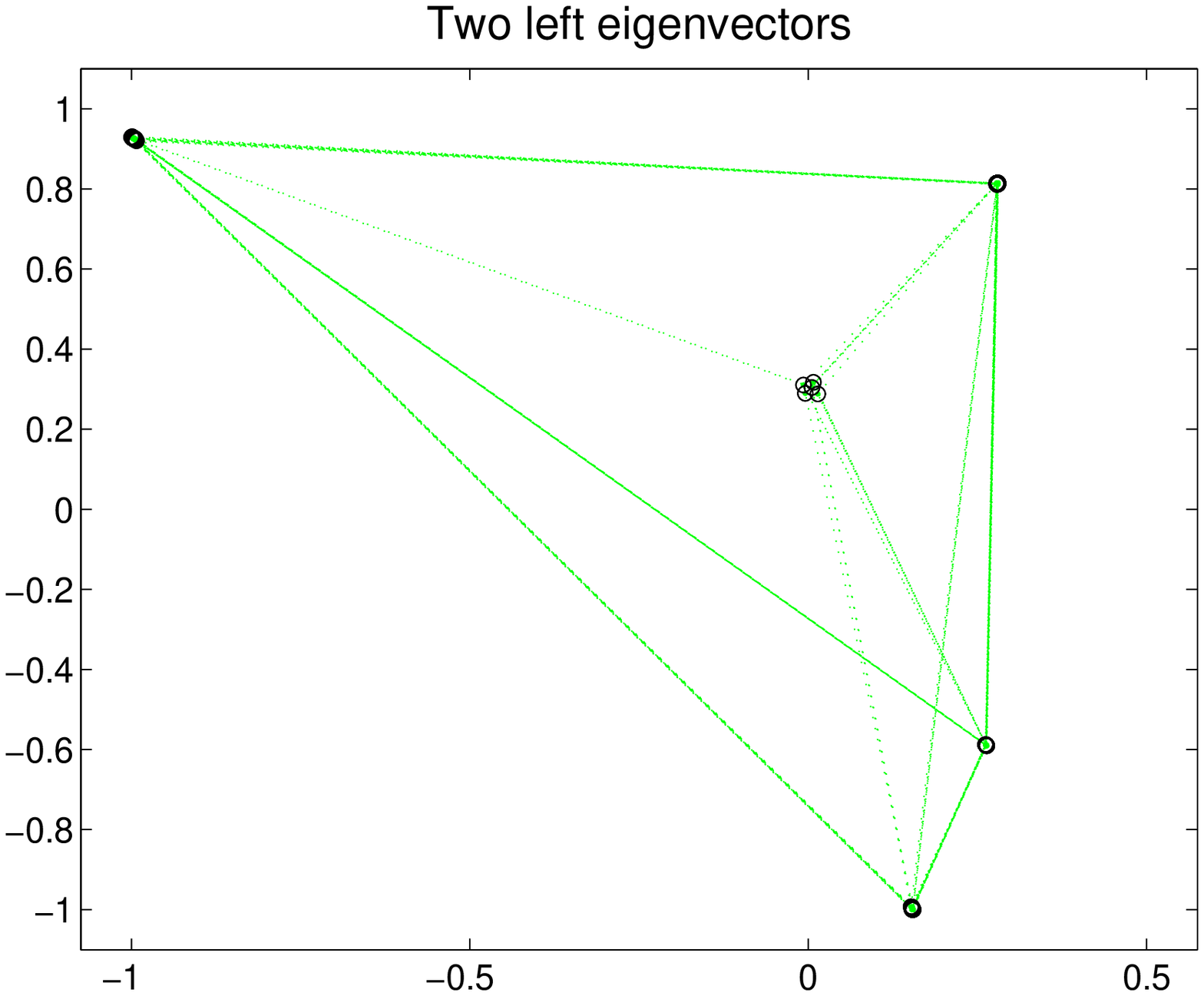}
\caption{(Color online) For the 4-phase case, if one plots only in 2 dimensions one does not see only 3 extremal points. The fourth apparently sticks out of the triangle formed by the others (although it did not have to) and in a third dimension actually forms a node of a tetrahedron (Fig.\ \ref{fig:extremals4}).}
\label{fig:extremals5}
\end{figure}

In the next figures we do the same exercise but with four phases, that is, \Eqref{eq:3phaseW} is modified by putting in a fourth block, $W_4$. Fig.\ \ref{fig:extremals3} shows the real part of the spectrum of $W$. The spectrum is similar to the 3-phase case, and as can be seen from the numbers, one does not require extremes of magnitude, large or small, to get useful information from the geometrical construction. By construction, the ``$W$'' of Fig.\ \ref{fig:extremals3} has three eigenvalues near the stationary one (0), leading to four phases. Fig.\ \ref{fig:extremals4} shows the convex hull of the points $\Abf(x)$. If one cuts off the plot in too low a dimension one gets what is seen in Fig.\ \ref{fig:extremals5}. Here only $A_1$ and $A_2$ are plotted and as can be seen there are four rather than three extrema. This is an illustration our need to have $\lambda_{m+1}$ much smaller than those preceding it. If this condition is not satisfied, more extrema appear. This is what Sec.\ \ref{sec:uniquenessproof} was all about.

\subsection{Hierarchical phases, no sharp cutoff in eigenvalue; simplified spin glasses\label{sec:spinglass}}

For two classes of phenomena we do not expect the eigenvalues to drop off suddenly, as discussed in connection with first order phase transitions. For spin glasses there is expected to be a hierarchical sequence of metastable states. For critical points the eigenvalues should have a power law dropoff near the stationary state.

\begin{figure}
\includegraphics[height=.4\textheight,width=.7\textwidth]{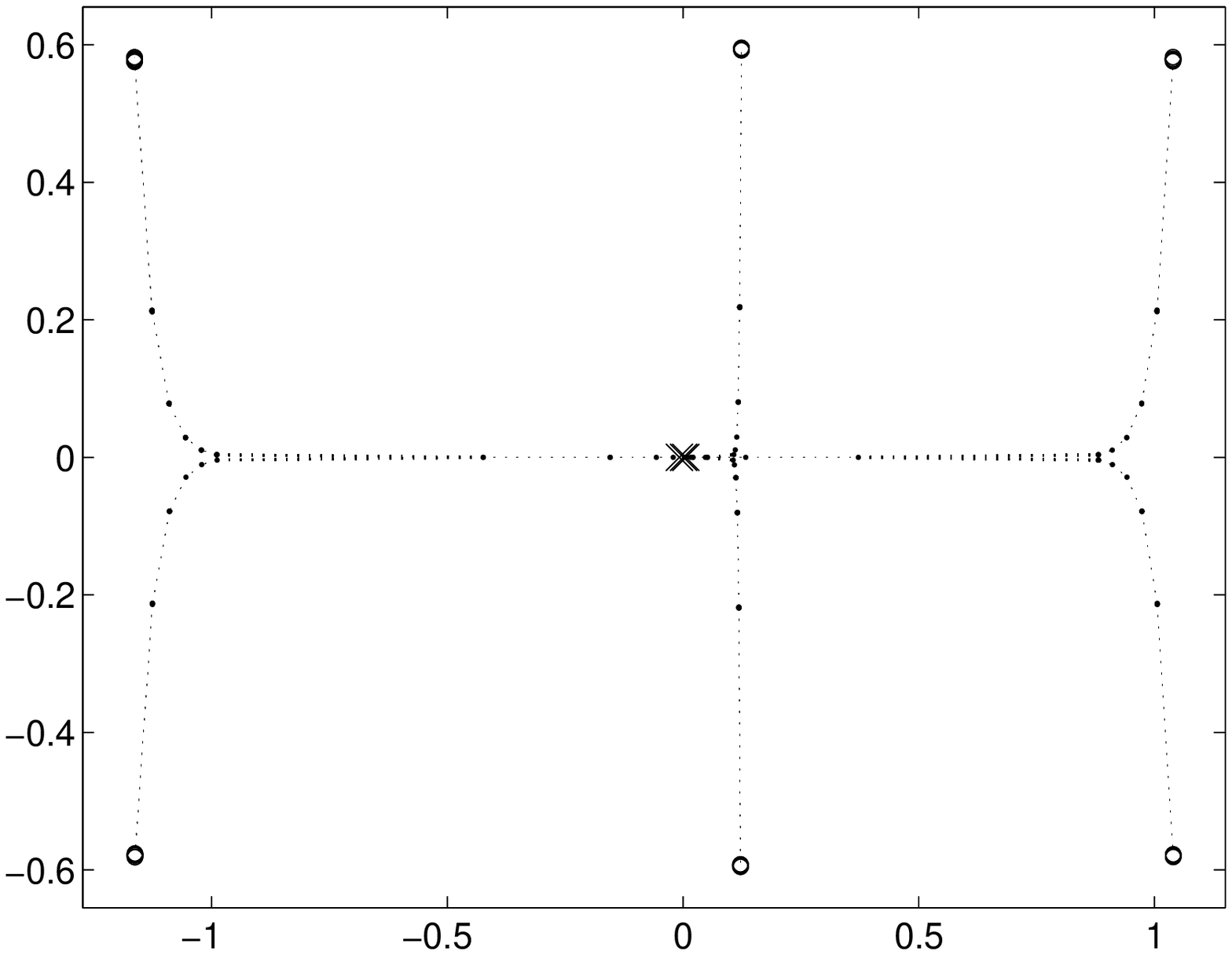}
\caption{Phases at successively later times for a hierarchical stochastic matrix. As explained in the text this is a two-dimensional projection of the five dimensional plot of eigenvectors multiplied by eigenvalue to the power $t$. On the shortest time scale there are 6 metastable phases (circles); subsequently they merge into three and finally into a single stationary state. The axes represent particular linear combinations of eigenvectors and lack physical dimensions (but have a scale determined by the normalization of Sec.\ \ref{sec:additionalproperties}). \label{fig:hierarchunfold}}
\end{figure}

For hierarchical structures, already studied by us in \cite{hierarchical}, we do a variant of the geometrical construction just displayed. The overall $W$ matrix has the following form
\be 
W=\lsmatrix{ccc}{
W_1&\wee&\wee\\
\wee&W_2&\wee\\
\wee&\wee&W_3} \,,\quad \hbox{with each $W_k$ of the form~}
W_k=\lsmatrix{cc}{
w_1&\weewee\\
\weewee&w_2}\,, 
\label{eq:hierarchicalW}
\ee
and $\epsilon\ll\delta\ll1$. For this structure it is instructive to introduce \textit{time} into the picture. The vectors to be plotted are $\Abft{t}(x)\equiv (A_1(x)\lambda_1^t,\dots,A_m(x)\lambda_m^t)$. We have built the hierarchy to have 6 phases; on a medium time scale three pairs of them decay into a common branch, subsequent to which the three branches merge into a single trunk. Since we cannot image the 5-dimensional structure, we take the projection of this motion (as a function of time) on a particular plane. This is shown in Fig.\ \ref{fig:hierarchunfold}, where the circles represent the original phases and the `$\times$' is the final state, $(0,0)$.

\subsection{Asymptotic probabilities\label{sec:asymptoticprob}}

\begin{figure}
\includegraphics[height=.4\textheight,width=.7\textwidth]{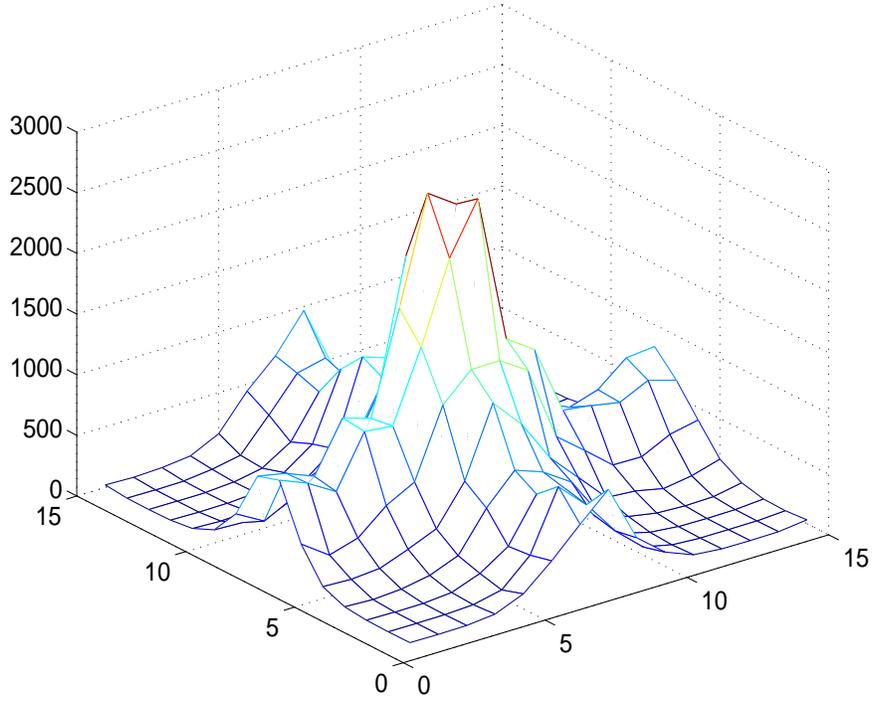}
\caption{(Color online) Landscape for a random walk on a 15 by 15 lattice. The distance unit on the lattice is arbitrary and the scale of the potential chosen so as to give a dynamical spectrum illustrating our representation. \label{fig:WalkPotential}}
\end{figure}

\begin{figure}
\includegraphics[height=.4\textheight,width=.7\textwidth]{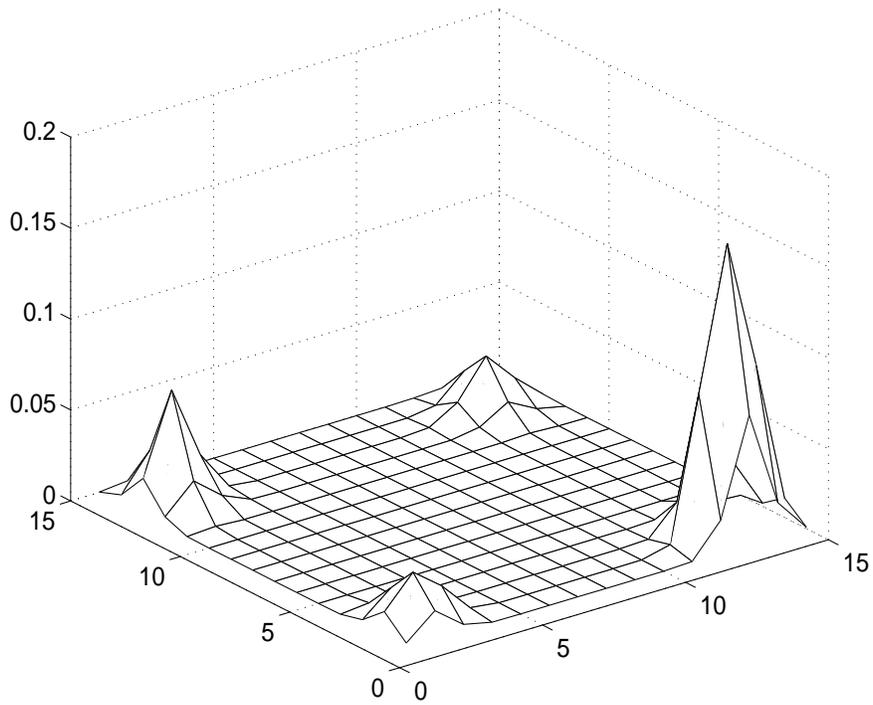}
\caption{Probability distribution for the stationary state of a walk on the landscape shown in Fig.~\ref{fig:WalkPotential}. \label{fig:WalkGroundState}}
\end{figure}

\begin{figure}
\includegraphics[height=.4\textheight,width=.7\textwidth]{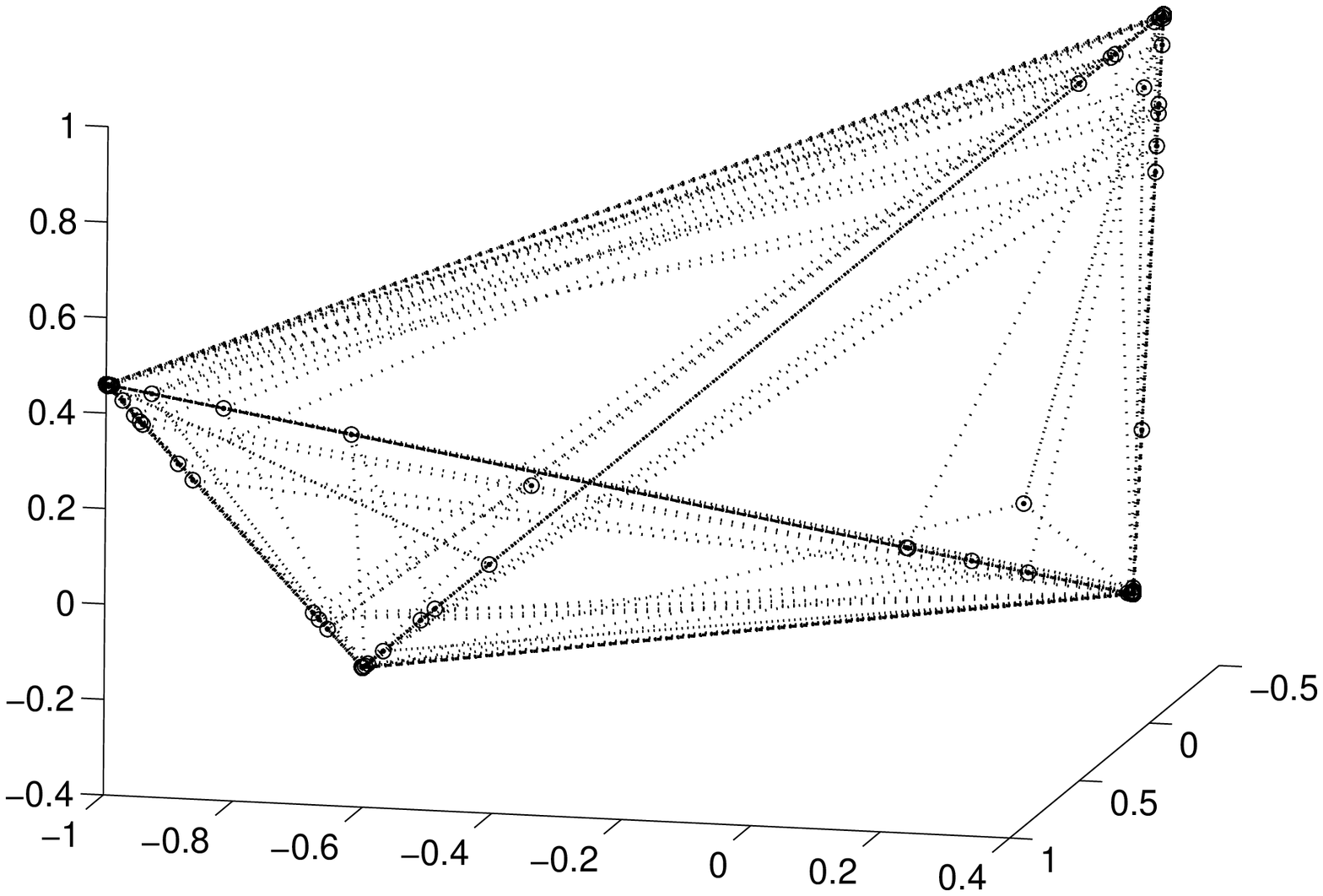}
\caption{Observable representation, in $\R^3$, of the states for the walk on the landscape shown in Fig.~\ref{fig:WalkPotential}. Each circle represents a point on the 15 by 15 lattice and its position within the tetrahedron (when expressed in barycentric coordinates with respect to the extremals) gives the probability of starting at that point and arriving at one or another extremal. The plot is very much like that shown in Fig.\ \ref{fig:extremals4}, but includes interior points. (Fig.\ \ref{fig:extremals4} shows only the convex hull.)  \label{fig:WalkBarycentric}}
\end{figure}

Consider a random walk on the landscape shown in Fig.\ \ref{fig:WalkPotential}. The stationary state is shown in Fig.\ \ref{fig:WalkGroundState}. There are clearly 4 regions of attraction, which we identify as the ``phases'' discussed in this article. The spectrum of the (225 by 225) generator of the stochastic dynamics is $[0,\exp(-16.0),\exp(-15.3),\exp(-14.8),\exp(+1.2),\dots]$, so that this satisfies the conditions for having 4 well-demarcated phases, which in this case represent regions of attraction. Finally in Fig.\ \ref{fig:WalkBarycentric} we show how the methods of this article can be used to calculate the probability that from a given initial condition one will arrive at one or another asymptotic state. Each circle in the graph (which is a 3-dimensional plot of a tetrahedron; cf. Fig.\ \ref{fig:extremals4}) represents a point on the 15 by 15 lattice and its location in the plot, when expressed in barycentric coordinates (positive numbers that add to 1) with respect to the extremals, gives its probability of reaching a particular phase. In the graph we do not identify the particular circles, but the same computer program that generated the graph can easily provide a table of probabilities for each initial condition. 

\section{Prospects\label{sec:prospects}}

The transition matrix for a stochastic process gives rise to \textit{observables}, namely its slowest left eigenvectors (in our convention $R_{xy}=\Pr\left(x\leftarrow y\right)$). For each $x$ in the state space of the process one can form a vector $\left(A_1(x),\dots,A_m(x)\right)$ for integer $m$, with $A_k$ the left eigenvectors. Depending on the spectrum of $R$, the space of these vectors can provide a graphic demonstration of the phases (in the sense of phase transitions) of this process. We call a plot of the points of the state space using a collection of slow (left) eigenvectors an \textit{observable-representation of state space}.

We have shown how when there is a hierarchical structure of phases that structure becomes manifest in the space of observables. Our model for demonstrating this is artificially constructed, but we expect that for systems of greater intrinsic interest the same features seen here should emerge. Thus, spin glass models could be considered, for example the Sherrington-Kirkpatrick model. Even for local spin glasses, although the state space grows large quite rapidly, our method requires little information from the transition matrix, $R$. For example, a 2-dimensional 4 by 4 spin glass would involve a 2$^{16}$ by 2$^{16}$ matrix (2$^{16}=65\,536$), but it is a \textit{sparse} matrix, and all we would want to know would be the first few eigenvectors, which is quite feasible. Now 4$\times$4 may not be much of a lattice, but the knowledge gained from the corresponding observables would immediately give information for the longest possible time scales. For the mean field Sherrington-Kirkpatrick model one should be able to do even better.

In the more traditional arena of stochastic processes, our geometric construction allows one to read off the probabilities of an initial point reaching any of various asymptotic states, even when one does not have prior knowledge of what those states are. We gave a simple example of a random walk on a multi-well landscape, but other examples easily come to mind. 

At the mathematical level, we believe our assumptions are stronger than they need to be. It is likely that hypothesis \ess\ can be replaced by something weaker. We already have preliminary results on this point in low dimension. Another place where we have an implicit assumption (although we did not emphasize it at the time) is in the proof of assertion (1) of the lemma, where we make a generic assumption about the geometry of the linear forms: specifically that angles in the effective coordinate system are not such that large values of $h$ could be generated from small distances. 

Cases where one has eigenvalues near one but there is not a sharp dropoff after one particular eigenvalue are of great importance in physical applications. Certainly for spin glasses, although they are expected to show the hierarchical structure discussed above, they would have a collection of time scales of decreasing size (local relaxation times) with no cutoff at a particular value. Also of interest is the case of critical phenomena. Here, absent hierarchical structure, we do not expect a small number of extrema to dominate. One also has other properties, for example the divergence of spatial correlation lengths, that on the face of it do not appear to be directly related to the dynamics. Nevertheless, as shown in \cite{grains}, much of this structure can be recovered from the eigenvectors, so that a dynamical characterization of this kind of transition, as we have done here for first order transitions, may well be possible.

\begin{figure}
\centerline{\includegraphics[height=3truein,width=3truein]{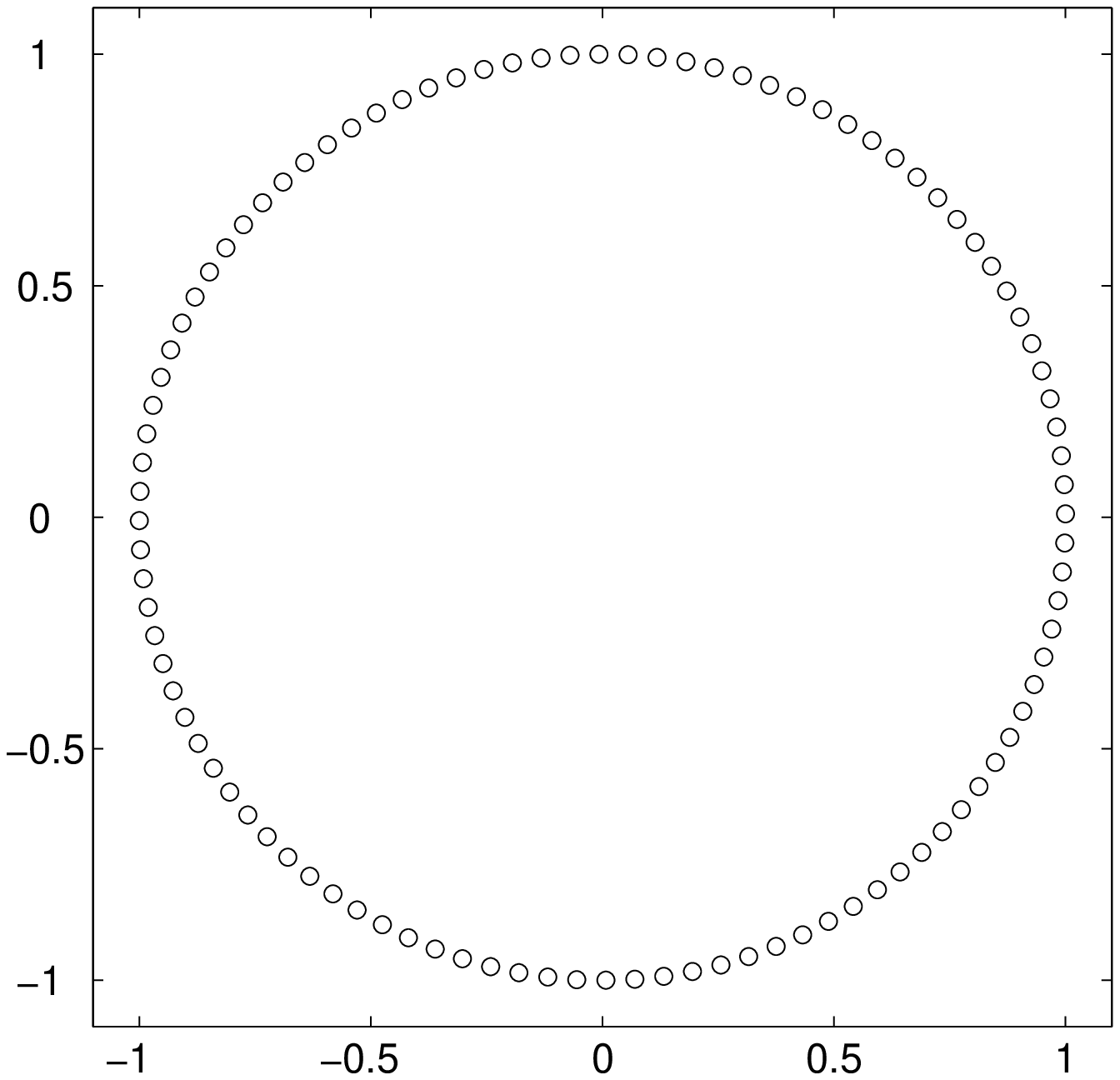}
\includegraphics[height=3.055truein,width=3.055truein]{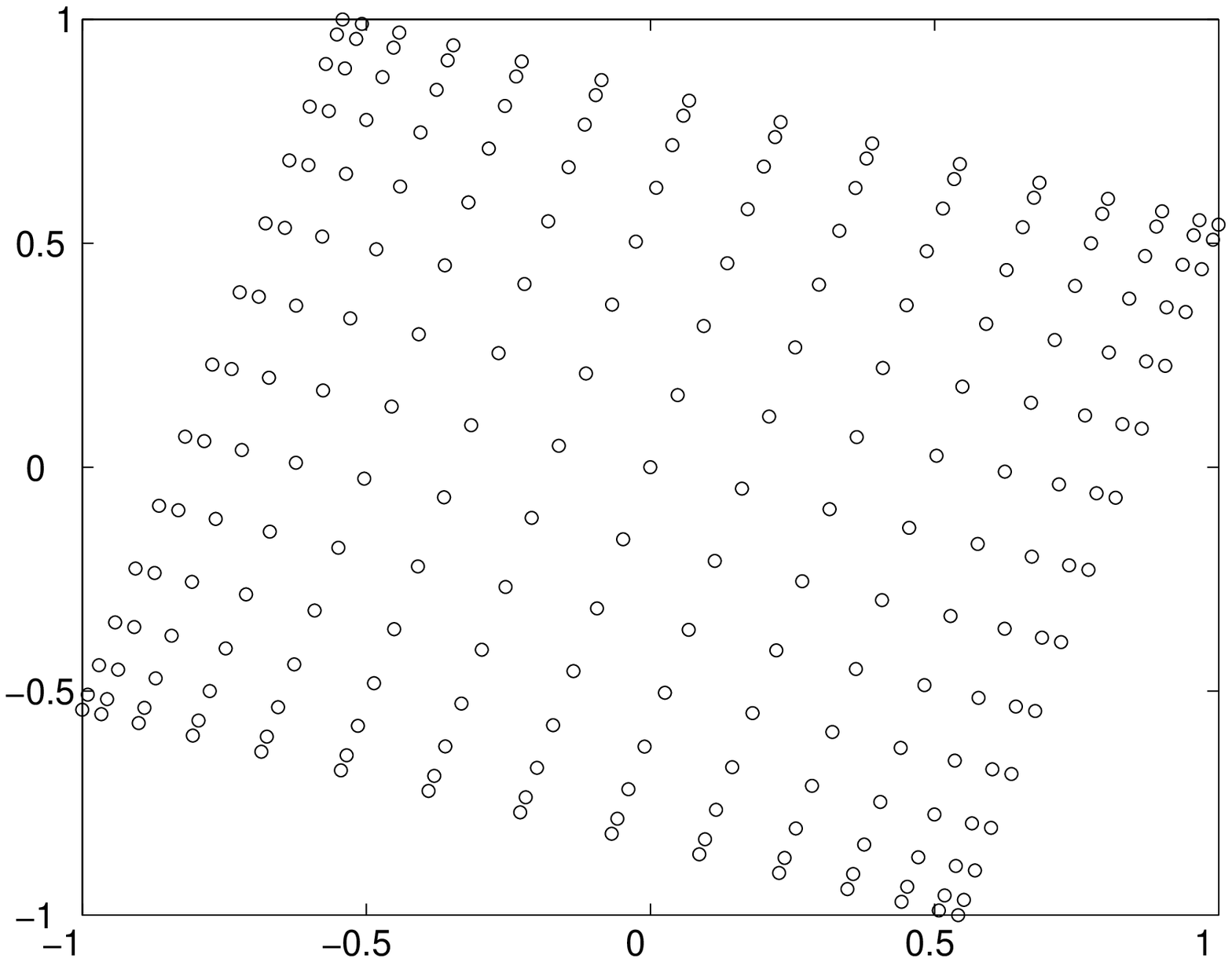}}
\caption{Observable representation, in $\R^2$, of the states for Brownian motion. The circle is for the one dimensional case of a walk on a ring. The rectangular figure is for a two-dimensional random walk with non-periodic boundary conditions.  \label{fig:brownianmotion}}
\end{figure}

Nevertheless, even where there is no phase transition, the observable-representation can provide an image of the state space. For the case of Brownian motion on a ring, the transition matrix simply has constants just off the diagonal, as well as in the corners to provide periodicity. We showed in \cite{grains} how $R$ can recover spatial structure, but a plot of $A_1(x)$ vs.\ $A_2(x)$, as in Fig.\ \ref{fig:brownianmotion}, is even more direct. As can be seen, this immediately gives the coordinate space ring. For two dimensions we show (in the same figure) a slight variant in which we have reflecting rather than periodic boundary conditions.

\section*{Acknowledgments} 
 This work was supported by NSF grant PHY 00 99471. LSS thanks the University of Paris VI and the Max Planck Institute for the Physics of Complex Systems, Dresden, for their kind hospitality.


\begin{thebibliography}{17}
\bibitem[{not({\natexlab{a}})}]{note:dustjordan}
\bibinfo{note}{In \Eqref{eq:dust}, $B^{(t)}$ is represented by a spectral
  decomposition. This is for convenience only. No harm would come if a Jordan
  form would be needed; all we require is the smallness of $B^{(t)}$.}
\bibitem[{not({\natexlab{b}})}]{note:qexplanation}
\bibinfo{note}{Note that the $q$ in \Eqref{eq:qdef2} corresponds to
  $y=y_\ell^*$ for some $\ell$, while in \Eqref{eq:qdef} it can refer to any
  point in $X$.}
\bibitem[{not({\natexlab{d}})}]{note:heuristic}
\bibinfo{note}{The present discussion is at a heuristic level, since in
  one-dimension, by the analyticity definition, there is no phase transition.
  However, the features observed here should persist in two dimensions and
  indeed in any phase transition of this sort. Note too, that for us metastable
  phases stand on an equal footing with stable ones and that our formalism
  abandons the thermodynamic limit. Rather, as should be evident here and in
  \cite{firstorder}, a phase transition is an asymptotic concept, so that we
  have no difficulty speaking of a phase transition in the one-dimensional
  model.}
\end{thebibliography}
\end{document}